\documentclass[preprint]{aastex}
\usepackage{epsf}

\shorttitle{Quasar Clustering and Lifetimes}
\shortauthors{Martini \& Weinberg}

\begin{document}

\newcommand{\mmin}{M_{\rm min}}
\newcommand{\beff}{b_{\rm eff}}
\newcommand{\hmpc}{h^{-1}{\rm Mpc}}
\newcommand{\tq}{t_Q}
\newcommand{\tu}{t_U}
\newcommand{\om}{\Omega_M}
\newcommand{\ol}{\Omega_\Lambda}
\newcommand{\ob}{\Omega_b}
\newcommand{\neff}{n_{\rm eff}}
\newcommand{\ronem}{r_{1m}}
\newcommand{\ltq}{{\rm log}_{10} \, t_Q}

\slugcomment{Accepted for publication in {\it The Astrophysical Journal} }

\title{Quasar Clustering and the Lifetime of Quasars}

\author{Paul Martini\altaffilmark{1} \& David H. Weinberg}

\affil{Department of Astronomy, 140 W. 18th Ave., Ohio State University, \\
Columbus, OH 43210 \\
martini@ociw.edu,dhw@astronomy.ohio-state.edu}

\altaffiltext{1}{Current Address: Carnegie Observatories, 813 Santa Barbara St., Pasadena, CA 91101}

\begin{abstract}

Although the population of luminous quasars rises and falls over
a period of $\sim~10^9$ years, the typical lifetime of individual quasars
is uncertain by several orders of magnitude.  We show that quasar
clustering measurements can substantially narrow the range of
possible lifetimes with the assumption that luminous quasars reside
in the most massive host halos.  If quasars are long-lived, then they
are rare phenomena that are highly biased with respect to the underlying
dark matter, while if they are short-lived they reside in more typical
halos that are less strongly clustered.  For a given quasar lifetime,
we calculate the minimum host halo mass by matching
the observed space density of quasars, 
using the Press-Schechter approximation.
We use the results of Mo \& White to calculate the clustering of
these halos, and hence of the quasars they contain,
as a function of quasar lifetime. 
A lifetime of $\tq = 4 \times 10^7$ years, the $e$-folding timescale of 
an Eddington luminosity black hole with accretion efficiency $\epsilon = 0.1$, 
corresponds to a quasar correlation length $r_0 \approx 10 \hmpc$ in 
low-density cosmological models at $z = 2 - 3$; this value is consistent 
with current clustering measurements, but these have large uncertainties. 
High-precision clustering measurements from the 2dF and Sloan quasar surveys 
will test our key assumption of a tight correlation between quasar luminosity
and host halo mass, and if this assumption holds then they should determine 
$\tq$ to a factor of three or better. An accurate determination of the quasar 
lifetime will show whether supermassive black holes acquire most of their 
mass during high-luminosity accretion, and it will show whether the 
black holes in the nuclei of typical nearby galaxies were once the central 
engines of high-luminosity quasars. 

\end{abstract}

\keywords{galaxies: quasars -- cosmology: dark matter, large-scale structure of the universe}

\section{Introduction}

Mounting evidence for the existence of supermassive black holes in the
centers of nearby galaxies 
\citep[recently reviewed by, e.g.,][]{richstone98}
supports the long-standing hypothesis that quasars are powered by
black hole accretion \citep[e.g.,][]{salpeter64,zeldovich64,lyndenbell69}.
However, one of the most basic properties of quasars, the typical
quasar lifetime $\tq$, remains uncertain by orders of magnitude.
The physics of gravitational accretion and radiation pressure provides
one natural timescale, the $e$-folding time 
$t_e=M_{\rm BH}/\dot{M} = 4 \times 10^8\, \epsilon\, l$ years of 
a black hole accreting mass with a radiative efficiency
$\epsilon=L/\dot{M}c^2$ and shining at a fraction $l=L/L_E$ of
its Eddington luminosity \citep{salpeter64}.
But while $\epsilon \sim 0.1$ and $l \sim 1$ are plausible 
values for a quasar, it is possible that black holes accrete 
much of their mass while radiating at much lower efficiency,
or at a small fraction of $L_E$.
The task of determining $\tq$ must therefore be approached empirically.

The observed evolution of the quasar luminosity function imposes 
a strong upper limit on $\tq$ of about $10^9$ years, since the 
whole quasar population rises and falls over roughly this interval 
\citep[see, e.g.,][]{osmer98}.
The lifetime of individual quasars could be much shorter than the lifetime
of the quasar population, however, and lower limits of $\tq \sim 10^5$ years
rest on indirect arguments, such as the requirement that quasars maintain
their ionizing luminosity long enough to explain the proximity effect
in the Ly$\alpha$ forest \citep[e.g.,][]{bajtlik88,bechtold94}.
A typical lifetime $\tq \sim 10^9$ years would imply that quasars are
rare phenomena, arising in at most a small fraction of high-redshift
galaxies.  Conversely, a lifetime as low as $\tq \sim 10^5$ years would imply
that quasars are quite common, suggesting that a large fraction of
present-day galaxies went through a brief quasar phase in their youth.

The comoving space density $\Phi(z)$ of active quasars at redshift $z$
is proportional to $\tq n_H(z)$, where $n_H$ is the comoving space density
of quasar hosts.  ``Demographic'' studies of the local black
hole population \citep[e.g.,][]{mag98,salucci99,marel99}
have opened up one route to determining the typical quasar lifetime:
counting the present-day descendants of the quasar central engines
in order to estimate $n_H(z)$ and thus constrain $\tq$ by matching $\Phi(z)$.
Roughly speaking, the ubiquity of black holes in nearby galaxies
suggests that quasars are common and that $\tq$ is likely in the 
range $10^6$ -- $10^7$ \citep[e.g.,][]{richstone98,haehnelt98,salucci99}.
However, as \citet{richstone98}
emphasize, the lifetime estimated in this way depends crucially on 
the way one links the mass of a present-day black hole to the 
luminosity of a high-redshift quasar, which in turn depends on assumptions
about the growth of black hole masses since the quasar epoch via
mergers or low-efficiency accretion.

In this paper we propose an alternative route to the quasar lifetime,
using measurements of high-redshift quasar clustering.
The underlying idea goes back to the work of \citet{kaiser84}
and \citet{bbks86}:
in models of structure formation based on 
gravitational instability of Gaussian primordial fluctuations, the
rare, massive objects are highly biased tracers of the underlying
mass distribution, while more common objects are less strongly biased.
Therefore, a longer quasar lifetime $\tq$ should imply a more clustered
quasar population, provided that luminous quasars reside in massive hosts.
The specific calculations that we present in this paper use the
Press-Schechter (1974; hereafter PS) 
approximation for the mass function of dark matter
halos and the \citet[hereafter MW]{mw96} and \citet{jing98} 
approximations for the bias of these
halos as a function of mass.  The path from clustering to quasar lifetime
has its own uncertainties; in particular, our predictions for quasar
clustering will rely on the assumption that the luminosity of a quasar
during its active phase is a monotonically increasing function of the
mass of its host dark matter halo.  However, the assumptions in the
clustering approach are at least very different from those in the black
hole mass function approach, and they can be tested empirically by
detailed studies of quasar clustering as a function of luminosity and
redshift. 

Our theoretical model of quasar clustering follows a general trend
in which the study of quasar activity is embedded in the broader 
context of galaxy formation and gravitational growth of structure
\citep[e.g.,][]{efstathiou88,turner91,haehnelt93,katz94,haehnelt98,haiman98,
monaco00,kauffmann00}.
This paper also continues a theme that is prominent in recent work
on the clustering of Lyman-break galaxies, namely that the clustering
of high-redshift objects is a good tool for understanding the physics
of their formation and evolution 
\citep[e.g.,][]{adelberger98,katz99,kolatt99,mmw99}.
Our model of the quasar population is idealized, but by focusing
on a simple calculation with clearly defined predictions, we hope to
highlight the link between quasar lifetime and clustering strength.
After presenting the theoretical results, we will draw some inferences
from existing estimates of the quasar correlation length.
However, our study is motivated mainly by the anticipation of vastly
improved measurements of quasar clustering from the 2dF 
and Sloan quasar surveys \citep[see, e.g.,][]{boyle99,fan99,york00}.
These measurements can test various
hypotheses about the origin of quasar activity, including our primary
assumption of a monotonic relation between quasar luminosity and host 
halo mass.  If this assumption proves valid, then the first major
physical result to emerge from the 2dF and Sloan measurements of high-redshift
quasar clustering will be a new determination of the typical quasar
lifetime.

\section{Method} \label{sec:meth}

\subsection{Overview} \label{sec:over}

We adopt a simple model of the high-redshift quasar population that is,
doubtless, idealized, but which should be reasonably accurate for our purpose
of computing clustering strength as a function of quasar lifetime.
We assume that all quasars reside in dark matter halos and that
a given halo hosts at most one active quasar at a time.
The first assumption is highly probable, since a dark matter collapse
is necessary to seed the growth of a black hole, and the second
should be a fair approximation at high redshift, where the masses
of large halos are comparable to the masses of individual galaxy
halos today.

Our strongest and most important assumption is that the luminosity of
a quasar during its active phase is monotonically related to the mass
of its host dark matter halo, and that all sufficiently massive halos
host an active quasar at some point.  More precisely, we assume 
that an absolute-magnitude limited sample of quasars at redshift $z$
samples the most massive halos present at that redshift,
and that the probability that a halo above the minimum host mass $\mmin$
harbors an active quasar at any given time is $\tq/t_H$, where
$\tq$ is the average quasar lifetime and $t_H$ is the halo lifetime.
We can therefore compute the value of $\mmin$ for a quasar population
with comoving space density $\Phi(z)$ from the condition
\begin{equation}
\Phi(z) = \int_{M_{\rm min}}^\infty dM n(M) {\tq \over t_H} .
\label{eqn:phimatch}
\end{equation}
We compute $n(M)$ using the PS approximation, and we compute the
bias of halos with $M>\mmin$ using the MW approximation.

A connection between quasar luminosity and host halo mass is
plausible on theoretical grounds --- the cores of massive halos collapse 
early, giving black holes time to grow, and these halos provide larger gas
supplies for fueling activity.  It is also plausible on empirical
grounds --- local black hole masses are correlated with the host
spheroid luminosity \citep{mag98,marel99,salucci99},
which in turn is correlated with stellar velocity dispersion 
\citep{faber76}.
A precisely monotonic relation is certainly an idealization, and we
explore the effects of relaxing this assumption in \S \ref{sec:sen}. 
The assumption of an approximately monotonic relation can be tested
empirically by searching for the predicted relation between clustering
strength and luminosity, as we discuss in \S \ref{sec:pros}.

The ubiquity of black holes in luminous local spheroids supports our
assumption that all sufficiently massive halos go through a quasar phase.
However, once the quasar space density declines at $z<2$, the occurrence
of quasar activity must be determined by fueling rather than by
the mere existence of a massive black hole, so it is not plausible that all
large halos host a {\it low-redshift} quasar.  We therefore apply
our model only to the high-redshift quasar population, at $z \geq 2$.

We implicitly assume that a quasar turns on at a random point in the
life of its host halo.  In this sense, our model differs subtly from
that of \citet{haehnelt98},
who assume that a quasar turns on when
the halo is formed, but this difference is unlikely to have a significant
effect on the predicted clustering.  
\citet{haehnelt98} pointed out that a longer quasar lifetime
would correspond to stronger quasar clustering because of the 
association with rarer peaks of the mass distribution, but they
did not calculate this relation in detail.

Because the quasar lifetime enters
our calculation only through the probability $\tq/t_H$ that a halo
hosts an active quasar at a given time, it makes no difference whether
the quasar shines continuously or turns on and off repeatedly with a
short duty cycle (as argued recently by \citealt{ciotti99}).
For our purposes, $\tq$ is the total time that the
quasar shines at close to its peak luminosity.
We also assume that quasars radiate isotropically, with a beaming
factor $f_B=1$, but because a smaller beaming factor simply changes
the conversion between observed surface density and intrinsic comoving space
density, all of our results can be scaled to smaller average beaming
factors by replacing $\tq$ with $f_B \tq$.

\subsection{Notation} \label{sec:note}

All of our calculations assume Gaussian primordial fluctuations.
We denote by $P(k)$ the power spectrum of these fluctuations
as extrapolated to the present day ($z=0$) by linear theory.
The rms fluctuation of the linear density field on mass scale $M$ is
\begin{equation}
\sigma(M) = \left[\frac{1}{2 \pi^2} \int_0^{\infty} dk \; k^2 \; P(k) 
\widetilde{W}^2(kr)\right]^{1/2}~,
\label{eqn:sigma}
\end{equation} 
where
\begin{equation}
\widetilde{W}(kr) = {3(kr\sin kr - \cos kr) \over (kr)^3}, \qquad
  r = \left(3M \over 4\pi\rho_0\right)^{1/3}
\label{eqn:window}
\end{equation}
is the Fourier transform of a spherical top hat containing average
mass $M$.  The mean density of the universe at $z=0$ is
$\rho_0 = 2.78\times 10^{11}\om h^2 \;M_\odot\;{\rm Mpc}^{-3}$,
with $h \equiv H_0/(100\;{\rm km}\;{\rm s}^{-1}\;{\rm Mpc}^{-1})$.
The rms fluctuation can be considered as a function of either 
the mass scale $M$ or the equivalent radius $r$.
We define the normalization of the power spectrum by the value of 
$\sigma_8 \equiv \sigma(r=8\hmpc)$.

The rms fluctuation of the linear density field at redshift $z$ is
\begin{equation}
\sigma(M,z) = \sigma(M)D(z),
\label{eqn:sigmaz}
\end{equation}
where $D(z)$ is 
the linear growth factor $D(z)$, defined so that $D(z=0)=1$.
The general expression for the growth factor in terms of the scaled
expansion factor $y=(1+z)^{-1}$ is 
\begin{equation}
\delta(y) = \frac{5}{2}\om {1 \over y} {dy \over d\tau}
       \int_0^y \left({dy^\prime \over d\tau}\right)^{-3} dy^\prime ,
\label{eqn:gfac}
\end{equation}
where $D(y) = \delta(y)$ for $\om = 1$, $D(y) = \delta(y)/\delta(1)$ 
for $\om < 1$, and the dimensionless time variable is $\tau=H_0 t$
\citep{heath77,carroll92}.
If the dominant energy components are pressureless matter and a
cosmological constant with $\Omega_\Lambda=\Lambda/3H_0^2$, then
the Friedmann equation implies
\begin{equation}
\left({dy \over d\tau}\right)^2 = 
    1 + \Omega_M\left(y^{-1} - 1\right) + \Omega_\Lambda\left(y^2-1\right).
\label{eqn:dydt}
\end{equation}
For an $\om = 1$, $\ol=0$ universe, $D(z)=(1+z)^{-1}$.
\citet[eq.~11.16]{peebles80} gives an exact analytic expression for
$D(z)$ for the case $\om<1$, $\ol=0$,  
and \citet[eq.~29]{carroll92} give an accurate analytic approximation for
$\Omega_\Lambda \neq 0$.
In our notation, $\sigma(M)$ without any explicit $z$ always refers
to the rms linear mass fluctuation on scale $M$ at $z=0$.

At any redshift, we can define a characteristic mass $M_*(z)$
by the condition
\begin{equation}
\sigma\left[M_*(z)\right] = \delta_c(z) = {\delta_{c,0} \over D(z)},
\label{eqn:mstar}
\end{equation}
where $\delta_c(z)$ is the threshold density for collapse of a 
homogeneous spherical perturbation at redshift $z$.
Because we implicitly define the density field as ``existing'' at
$z=0$, the collapse threshold $\delta_c(z)$ increases with increasing
redshift.  For an $\om=1$, $\ol=0$ universe,
$\delta_{c,0} = 0.15(12\pi)^{2/3} \approx 1.69$
(see, e.g., \citealt{peebles80}, \S 19).
For other models, we incorporate the dependence of $\delta_{c,0}$ on
$\om$ in Appendix A of \citet{nfw97}, but
because $\om$ approaches one at high redshift in all models
this correction to $\delta_c$ is less than 2\% in all of the cases that 
we consider.

\subsection{From the Quasar Lifetime to the Minimum Halo Mass} \label{sec:life}

For a specified quasar lifetime, we compute the minimum halo mass
by matching the comoving number density $\Phi(z)$ of observed quasars,
accounting for the fact that only a fraction $\tq/t_H$ of host halos
will have an active quasar at the time of observation.
The matching condition is equation~(\ref{eqn:phimatch}), or,
putting in explicit mass and redshift dependences,
\begin{equation}
\Phi(z) = \int_{M_{\rm min}}^\infty dM {\tq \over t_H(M,z)} n(M,z).
\label{eqn:phimatch2}
\end{equation}
If $\tq>t_H(M,z)$, we set the factor $\tq/t_H$ to unity.
For the halo number density we use the PS approximation,
\begin{equation}
n(M, z)\;dM = - \sqrt{\frac{2}{\pi}} \frac{\rho_0}{M} 
		\frac{\delta_c(z)}{\sigma^2(M)} \frac{d\sigma(M)}{dM} \; 
		{\rm exp} \left[ -\frac{\delta_c^2(z)}{2 \sigma^2(M)} \right]
		dM ,
\label{eqn:ps}
\end{equation}
where $\rho_0$ is the mean density of the universe at $z=0$,
$\sigma(M)$ is the rms fluctuation given by equation~(\ref{eqn:sigma}),
and $\delta_c(z)$ is the critical density for collapse by redshift $z$.

In a gravitational clustering model of structure formation,
halos are constantly growing by accretion and mergers, so the
definition of a ``halo lifetime'' is somewhat ambiguous.
For $\om(z) \approx 1$, a typical halo survives
for roughly a Hubble time
before being incorporated into a substantially larger halo,
since the age of the universe at redshift $z$
is also the characteristic dynamical time of objects forming at that redshift.
Thus, to a first approximation, one could simply substitute
$t_H(M,z)=\tu(z)$ in equation~(\ref{eqn:ps}).
We can do somewhat better by using the extended Press-Schechter
formalism 
\citep[e.g.,][]{bond91,lc93}
to calculate
the average halo lifetime, thereby accounting for the dependence
of $t_H$ on the power spectrum shape and the halo mass.
Structure grows more rapidly in a cosmology with a redder power
spectrum, and more massive halos accrete mass more rapidly.

Equation~(2.22) of \citet{lc93} gives the probability that
a halo of mass $M_1$ existing at time $t_1$ will have been incorporated
into a new halo of mass greater than $M_2$ by time $t_2$:
\begin{eqnarray}
P(S < S_2, \omega_2 | S_1, \omega_1) & = &\frac{1}{2} 
  \frac{(\omega_1 - 2 \omega_2)}{\omega_1} {\rm exp} 
  \left[ \frac{2 \omega_2 (\omega_1 - \omega_2)}{S_1}\right] 
  \left[ 1 - {\rm erf} \left( \frac{S_2 (\omega_1 - 2 \omega_2) + S_1 \omega_2}
  {\sqrt{2 S_1 S_2 (S_1 - S_2)}} \right) \right] \nonumber \\
  &&+ \frac{1}{2} \left[ 1 - {\rm erf} 
     \left( \frac{S_1 \omega_2 - S_2 \omega_1}{\sqrt{2 S_1 S_2 (S_1 - S_2)}} 
     \right) \right]~,
 \label{eqn:lc}
\end{eqnarray}
where $S_1 = \sigma^2(M_1)$, $S_2 = \sigma^2(M_2)$, $\omega_1 = \delta_c(t_1)$, 
and $\omega_2 = \delta_c(t_2)$. 
In this equation, $\omega$ plays the role of the ``time'' variable, with
$\omega_2<\omega_1$ corresponding to $t_2>t_1$, and $S$ plays the role
of the ``mass'' variable, with $S_2<S_1$ corresponding to $M_2>M_1$.
For a halo of mass $M$ existing at time $\tu(z)$, we define the halo lifetime
to be the median interval before such a halo is incorporated into a
halo of mass $2M$.  Thus, $t_H(M,z) = \hat{t}_S-\tu(z)$, where $\hat{t}_S$
is the time at which the probability in equation~(\ref{eqn:lc}) equals 0.5,
for $S_1=\sigma^2(M)$ and $S_2=\sigma^2(2M)$.
Clearly other plausible definitions of $t_H(M,z)$ are possible, 
and they would give answers different by factors of order unity.
With our definition, a black hole that lights up repeatedly 
is considered the ``same'' quasar as long as the mass of its host
halo remains the same within a factor of two.
If the host merges into a much larger halo and the black hole lights
up again, it is considered a ``new'' quasar.
We show the halo lifetimes for different masses and power spectra when we
discuss specific models below.

For comoving space densities $\Phi(z)$, we adopt values based on the 
work of \citet{boyle90}, \citet{hewett93}, and \citet{who94}.
Since observations constrain the number of objects per unit redshift
per unit solid angle, the conversion to comoving space density
depends on the values of the cosmological parameters.
We provide the formulas for these conversions in the Appendix, and in Table A1 
we list our adopted values of $\Phi(z)$ and the surface densities of objects 
to which these space densities correspond. In general, $\Phi(z)$
represents the space density of quasars above some absolute magnitude, 
corresponding to a surface density above some apparent magnitude. 
In \S \ref{sec:pros} we discuss how to scale our results to predict the 
clustering of samples with different measured surface densities. 

\subsection{From Minimum Halo Mass to Clustering Length} \label {sec:clust}

Halos with $M>M_*$ are clustered more strongly than the underlying
distribution of mass.  MW give an approximate formula,
\begin{equation}
b(M, z) = 1 + \frac{1}{\delta_{c,0}} 
	  \left[ \frac{\delta_c^2(z)}{\sigma^2(M)} - 1 \right],
\label{eqn:biasmw}
\end{equation}
for the bias factor of halos of mass $M$ at redshift $z$.
On large scales, the ratio of rms fluctuations in halo number
density to rms fluctuations in mass should be $b(M,z)$.
This formula is derived from an extended Press-Schechter analysis,
and it agrees fairly well with the results of N-body simulations
on scales where the rms mass fluctuations are less than unity.
The MW formula becomes less accurate for halos with $M<M_*$,
i.e., $\sigma(M)<\delta_c(z)$, and \citet{jing98} provides
an empirical correction that fits the N-body results,
\begin{equation}
b(M, z) = \left( 1 + \frac{1}{\delta_{c,0}} 
	  \left[ \frac{\delta_c^2(z)}{\sigma^2(M)} - 1 \right] \right)
          \left( \frac{\sigma^4(M)}{2\,\delta_c^4(z)} 
          + 1 \right)^{(0.06 - 0.02\neff)}, 
\label{eqn:bias}
\end{equation}
where $\neff=3-6\,(d\ln\sigma/d\ln M)$ 
is the effective index of the power spectrum on mass scale $M$.

According to our model, the quasars at redshift $z$ only reside in 
halos of mass $M>\mmin$.  The effective bias of these host halos
is the bias factor~(\ref{eqn:bias}) weighted by the number density
and lifetime of the corresponding halos:
\begin{equation}
\beff(\mmin,z) = \left(\int_{\mmin}^{\infty} dM\;
   \frac{b(M, z) n(M, z)}{t_H(M, z)}\right) 
    \left(\int_{\mmin}^{\infty} dM\;\frac{n(M, z)}{t_H(M, z)}\right)^{-1}.
\label{eqn:beff}
\end{equation}
Because the halo number density drops steeply with increasing mass,
the effective bias is usually only slightly larger than the bias
factor at the minimum halo mass, $b(\mmin,z)$.

As our measure of clustering amplitude, we use the radius $r_1$ of a top hat 
sphere in which the rms fluctuation $\sigma_Q$ of quasar number counts 
(in excess
of Poisson fluctuations) is unity.  This quantity is similar to the 
correlation length $r_0$ at which the quasar correlation function 
$\xi(r)$ is unity, but it can be
more robustly constrained observationally because it does not require
fitting the scale-dependence of $\xi(r)$.  For a power law correlation
function $\xi(r)=(r/r_0)^{-1.8}$, $r_1\approx 1.4 r_0$.
With our adopted approximation for the bias, $r_1$ is determined
implicitly by the condition
\begin{equation}
\sigma_Q(r_1,z) = \beff(\mmin, z)\sigma(r_1)D(z) = 1,
\label{eqn:r1def}
\end{equation}
where $\sigma(r_1)$ is the rms linear mass fluctuation at $z=0$ in
spheres of radius $r_1$.  For a specified cosmology, mass power spectrum
$P(k)$, quasar lifetime $\tq$, and comoving space density $\Phi(z)$,
we determine $r_1$ from equation~(\ref{eqn:r1def}),
computing $\sigma(r)$ 
from equation~(\ref{eqn:sigma}), $D(z)$ from equation~(\ref{eqn:gfac}),
$\mmin$ from equation~(\ref{eqn:phimatch2}), and $\beff(\mmin)$
from equations~(\ref{eqn:bias}) and~(\ref{eqn:beff}).

\subsection{Results for Power Law Power Spectra} \label{sec:plspec}

Models with a power law power spectrum, $P(k) \propto k^n$, provide
a useful illustration of our methods, since many steps of the calculation
can be done analytically.
For such models, the dependence of rms fluctuation on mass is also a
power law,
\begin{equation}
\sigma(M) ~=~ {\sigma(M,z) \over D(z)} ~=~ 
  {\delta_{c,0} \over D(z)} \left[{M \over M_*(z)}\right]^{-(3+n)/6} ~=~
  \delta_c(z) \left[{M \over M_*(z)}\right]^{-(3+n)/6},
\label{eqn:sigmapl}
\end{equation}
where $M_*(z)$ is the characteristic non-linear mass defined by
equation~(\ref{eqn:mstar}).
With this substitution, the PS mass function can be expressed as a
function of $M_*(z)$ and the dimensionless mass variable $x=M/M_*(z)$.
Integrating to obtain the comoving number density of objects
with mass $M>\mmin$ yields
\begin{equation}
N(M>\mmin) = \sqrt{\frac{2}{\pi}} \left(\frac{n+3}{6}\right) 
	     \left(\frac{\rho_0}{M_*}\right) \int^{\infty}_{x_{min}} dx\; 
	     x^{\frac{n-9}{6}} {\rm exp}\left[ -\frac{1}{2} 
	     x^{\frac{n+3}{3}} \right]. 
\label{eqn:psint}
\end{equation}

\begin{figure*}[t]
\centerline{
\epsfxsize=3.5truein
\epsfbox[65 165 550 730]{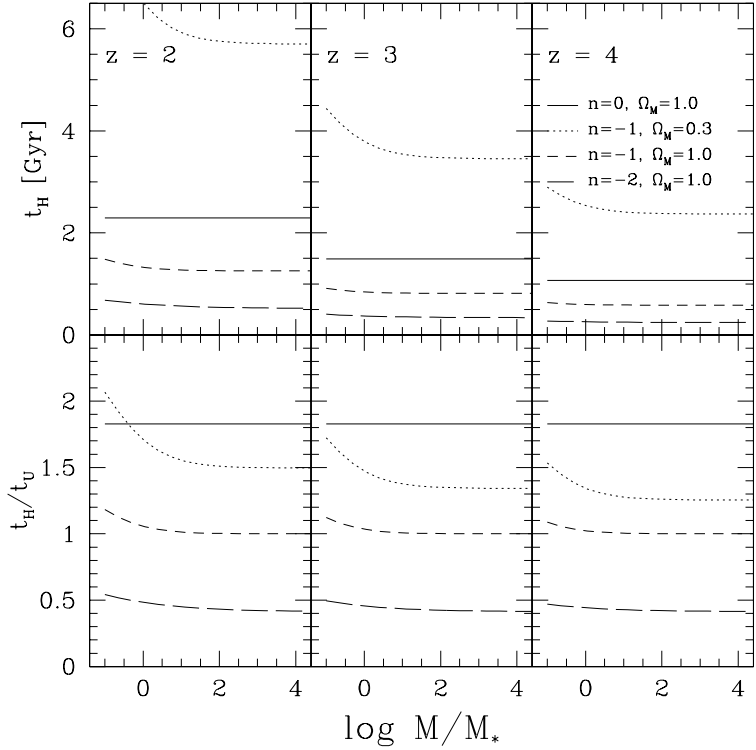}
}
\caption{
\footnotesize
Halo lifetimes vs. $M/M_*$ for the power law cosmologies listed in 
Table~\ref{tbl:pl}. Upper panels show the halo lifetimes in Gyr for 
each model at $z = 2$, 3, and 4. Lower panels show the ratio of the 
halo lifetime to the age of the universe. This ratio is independent of 
redshift for the $\om = 1$ models, but not for the $n = -1, \om = 0.3$ 
model.
} \label{fig:lifepl}
\end{figure*}

For power law models with $\om=1$, the ratio of the halo
lifetime $t_H(M,z)$ to the age of the universe at redshift $z$
depends only on $M/M_*(z)$ and has no separate dependence on redshift.
Figure~\ref{fig:lifepl} shows $t_H$ and the ratio $t_H/\tu$ as a function
of $M/M_*$ for power law models with $n=0$, $-1$, and $-2$.
More massive halos tend to accrete mass more quickly and therefore
have shorter median lifetimes.  At a given value of $M/M_*$, the
halo lifetime is shorter for lower $n$ because a greater amount
of large scale power causes the typical mass scale of non-linear
structure to grow more rapidly.  Although the calculation of the
median halo lifetime via equation~(\ref{eqn:lc}) is moderately
complicated, the median lifetime for large masses 
asymptotically approaches a constant value
\begin{equation}
t_H(M,z) = \left[2^{(3+n)/2}-1\right]\tu(z), \qquad
  M \gg M_*(z),~\Omega_M=1,
\label{eqn:tsimple}
\end{equation}
\citep{lc93}.
We will find below that the predicted masses
of quasar host halos are indeed in this asymptotic regime
for most plausible parameter choices.
The halo lifetime is longer for $\om<1$ than for $\om=1$
because fluctuations grow more slowly in a low-density universe, 
but $t_H$ still asymptotically approaches a constant value. 
The dotted curves 
in Figure~\ref{fig:lifepl} illustrate the case $n=-1$, $\om=0.3$,
$\ol=0$. The cosmological parameters for all of our models with power law
power spectra are summarized in Table~\ref{tbl:pl}. 

\begin{center}
\begin{deluxetable}{lrccclll}
\tablenum{1}
\tablewidth{0pt}
\tablecaption{Power Law Model Parameters \label{tbl:pl} }
\tablehead {
  \colhead{} &
  \colhead{} &
  \colhead{$\sigma_8$} &
  \colhead{$h$} &
  \colhead{$\om$} &
  \colhead{$M_* (z = 2)$} &
  \colhead{$M_* (z = 3)$} &
  \colhead{$M_* (z = 4)$} \\
}
\startdata
$n = 0$ &. . . . . . .     & 0.5  & 1.0  & 1.0 & $5.87\cdot 10^{12}$
  & $3.30\cdot 10^{12}$ & $2.11\cdot 10^{12}$    \\
$n = -1$ &. . . . . . .    & 0.5  & 1.0  & 1.0 & $5.80\cdot 10^{12}$
  & $2.45\cdot 10^{12}$ & $1.25\cdot 10^{12}$    \\
$n = -1$ &. . . . . . .    & 1.0  & 1.0  & 0.3 & $5.44\cdot 10^{12}$
  & $2.87\cdot 10^{12}$ & $1.70\cdot 10^{12}$    \\
$n = -2$ &. . . . . . .    & 0.5  & 1.0  & 1.0 & $5.59\cdot 10^{8}$
  & $9.96\cdot 10^7$    & $2.61\cdot 10^{7}$    \\
\enddata
\tablecomments{Parameters of the four power law cosmological models 
discussed in \S\ref{sec:plspec}. Columns 1 - 4 list the power spectrum 
index $n$, normalization $\sigma_8$, scaled Hubble constant $h$, and 
mass density parameters $\om$. 
Columns 5 - 7 list the values of $M_*$ (eq.~[\ref{eqn:mstar}])
at $z = 2$, 3, and 4, in units of $h^{-1} M_\odot$.
}
\end{deluxetable}
\end{center}

\begin{figure*}[t]
\centerline{
\epsfxsize=3.5truein
\epsfbox[65 165 550 730]{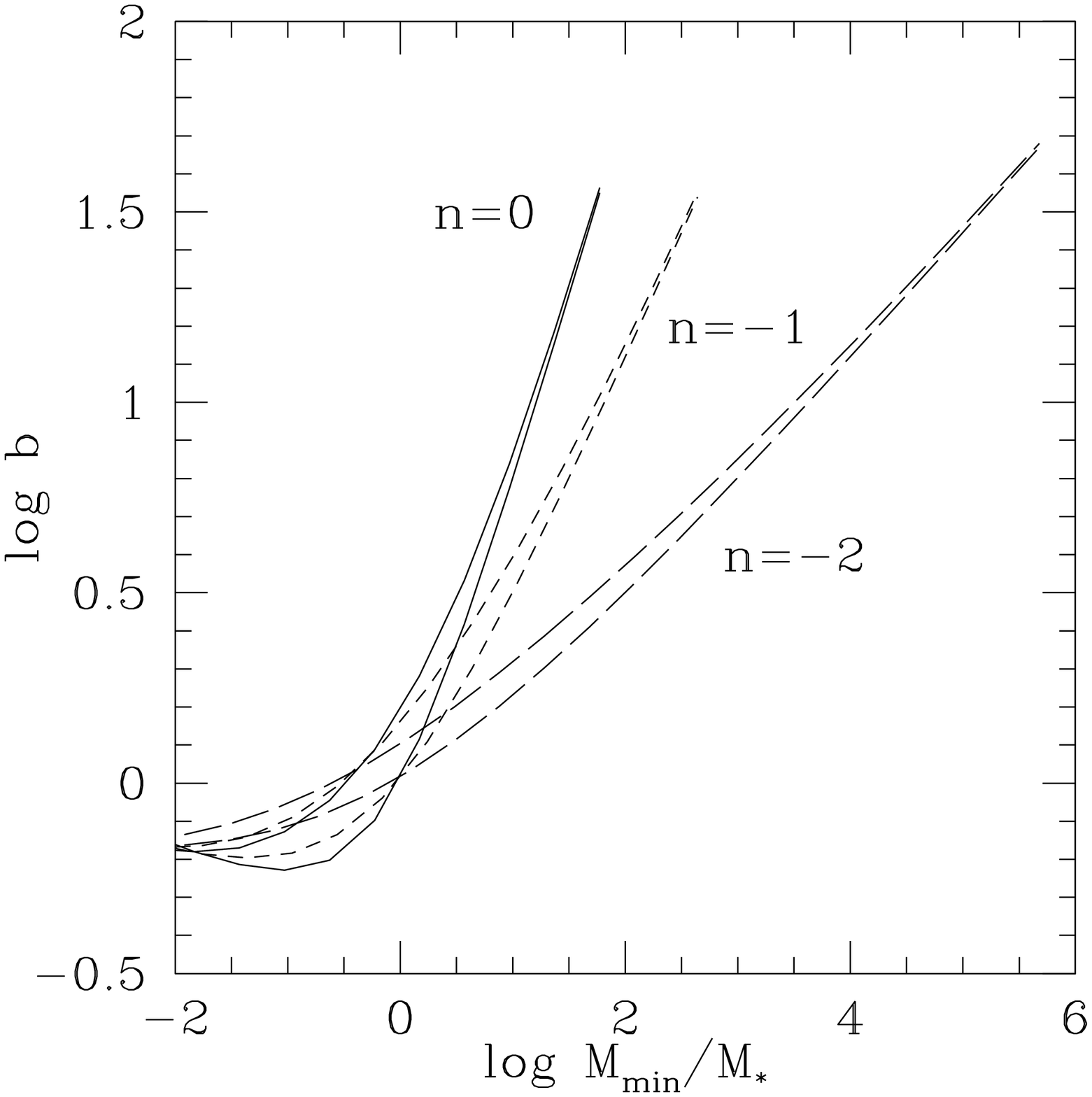}
}
\caption{
\footnotesize
Bias vs. $M/M_*$ for the power law models with $n = 0$ (solid), 
$-1$ (short-dashed), and $-2$ (long-dashed). Lower curves show $b(\mmin)$ 
computed from equation~(\ref{eqn:bias}), and upper curves show the 
number-weighted effective bias (eq.~[\ref{eqn:beff}]). 
} \label{fig:biaspl}
\end{figure*}

The power law scaling of the rms fluctuation amplitude,
equation~(\ref{eqn:sigmapl}), allows the bias
formula~(\ref{eqn:bias}) to be written
\begin{equation}
b(M,z) = \left(1 + {1 \over \delta_{c,0}}
         \left\{\left[{M \over M_*(z)}\right]^{(3+n)/3}-1\right\} \right) 
         \left( \frac{1}{2} \left[{M \over M_*(z)}\right]^{-2(3+n)/3}
         + 1 \right)^{(0.06 - 0.02n)}. 
\label{eqn:biaspl}
\end{equation}
Note that the second factor is very close to one for $M \geq M_*$. 
Figure~\ref{fig:biaspl} shows the bias and the corresponding
number-weighted effective bias (eq.~[\ref{eqn:beff}]) as a 
function of $\mmin/M_*$.  For $\mmin>M_*$, the 
effective bias is only slightly larger than $b(\mmin)$, 
since the number density of halos declines rapidly with
increasing $M$.  As equation~(\ref{eqn:biaspl}) shows,
the bias depends more strongly on $M$ for larger $n$.
However, the exponentially falling tail of the mass function
at high $M/M_*$ is much steeper for higher $n$, 
as one can see from equation~(\ref{eqn:psint}).
As a result, the bias at fixed comoving number density is
higher for {\it smaller} $n$ in the high $M/M_*$ regime 
(see Fig.~\ref{fig:r1pl} below). 

Under the (good) approximation that the halo lifetime is
given by the asymptotic formula~(\ref{eqn:tsimple}) in the
mass range of interest, the halo lifetime can be moved outside
of the integral~(\ref{eqn:phimatch2}) for the number density
of active quasars.  The implied quasar lifetime as a function of 
minimum halo mass is then
\begin{equation}
\tq(\mmin) = \frac{t_H \, \Phi(z)}{N(M>\mmin)}, 
\label{eqn:tqpl}
\end{equation}
where $N(M>\mmin)$ is given by equation~(\ref{eqn:psint}). For the 
$\om = 0.3, n = -1$ model, we also use 
the asymptotic value of $t_H$, though this is no longer given by 
equation~(\ref{eqn:tsimple}).
We use a $P(k)$ normalization $\sigma_8 = 0.5$ for the three
$\om=1$ models and $\sigma_8 = 1.0$ for the $\om=0.3$ model,
in approximate agreement with the constraint on $\sigma_8$ and $\om$
implied by the observed mass function of rich galaxy clusters
\citep{white93,eke96}.

Equation~(\ref{eqn:tqpl}) implicitly determines $\mmin/M_*(z)$ given $\tq$.
The top panels of Figure~\ref{fig:r1pl} show $\mmin/M_*(z)$ as a 
function of $\tq/\tu$ for $z = 2$, 3 and 4 and a constant 
comoving space density $\Phi(z)=10^{-6} h^3$ Mpc$^{-3}$. 
For the $\om=1$ cases, where $t_Q/t_U$ depends only on $n$ and
$\mmin/M_*$, the redshift dependence of $\mmin/M_*$ arises solely
from the presence of $M_*$ in the number density formula~(\ref{eqn:psint}).
As $M_*$ increases with decreasing redshift, the value of
$x_{\rm min}=\mmin/M_*$ must decrease to keep $\Phi(z)$ constant.
Smaller values of $n$ lead to higher values of $\mmin/M_*$ because
of the gentler fall off of the mass function at large $M/M_*$.  
The $n=-2$ curves become
flat at the largest values of $\tq/t_U$ because $\tq$ begins to exceed
the halo lifetime $t_H$, implying that all halos above $\mmin$ are
occupied by quasars.  The difference between the open and $\om=1$
curves for $n=-1$ reflects mainly the larger values of $\sigma_8$
and $D(z)$ in the open model, which lead to a lower value of $\rho_0/M_*$
in the mass function~(\ref{eqn:psint}) and therefore require a lower value 
of $\mmin/M_*$ to compensate.

\begin{figure*}[t]
\centerline{
\epsfxsize=3.5truein
\epsfbox[65 165 550 730]{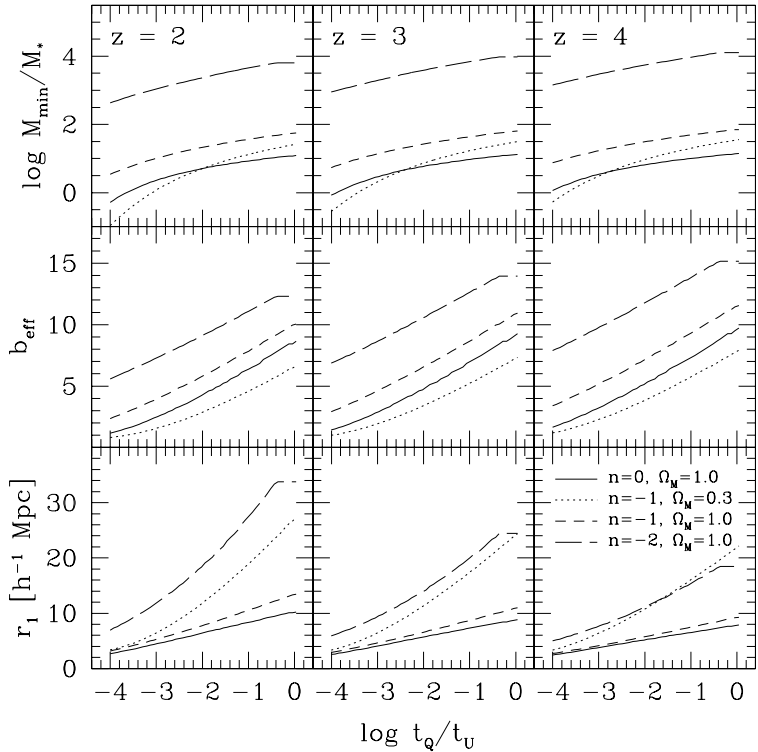}
}
\caption{
\footnotesize
Halo mass, bias, and clustering length for the power law models, 
as a function of quasar lifetime. Top panels show the minimum mass $\mmin$ 
required to obtain a space density $\Phi(z) = 10^{-6} h^3$ Mpc$^{-3}$ for 
a given value of $\tq/\tu$ at $z = 2$, 3 and 4. Middle panels show the 
corresponding effective bias $\beff$ for halos $\mmin$. 
Bottom panels show the clustering lengths $r_1$ as a function of $\tq/\tu$. 
The clustering length is the radius of a top hat sphere in which rms number 
count fluctuations (in excess of Poisson) are unity. For a power law 
correlation function $\xi(r) = (r/r_0)^{-1.8}$, $r_1 \approx 1.4 \, r_0$. 
} \label{fig:r1pl}
\end{figure*}

The middle panels of Figure~\ref{fig:r1pl} show the effective bias
$\beff(\mmin,z)$ for the power law models.  As already remarked, the
bias at fixed space density and $\tq/t_U$ is higher for redder power
spectra (smaller $n$) because of the much higher values of $\mmin/M_*$,
despite the partially counterbalancing effect of the stronger
dependence of bias on mass at larger $n$.  Physically, the higher
bias for redder spectra
reflects the greater influence of the large scale environment
on the amplitude of small scale fluctuations.
For a given model, the bias 
increases with increasing redshift, reflecting the increase in $\mmin/M_*$; 
the change, however, is quite modest.

The rms number count fluctuation on comoving scale $r$ is
\begin{equation}
\sigma_Q(r,z)=b_{\rm eff}(\mmin,z)\sigma(r,z)=
   b_{\rm eff}(\mmin,z) \sigma_8 D(z) \left({r \over 8\hmpc}\right)^{-(3+n)/2}.
\label{eqn:sigmaqpl}
\end{equation}
The quasar clustering length $r_1$ is the scale on which this rms
fluctuation amplitude is unity,
\begin{equation}
r_1 = 8\hmpc\,\times\, \left[b_{\rm eff}(\mmin,z)\sigma_8 D(z)\right]^{2/(3+n)}.
\label{eqn:r1pl}
\end{equation}
The bottom panels of Figure~\ref{fig:r1pl} present the main result of
this Section, the dependence of $r_1$ on 
quasar lifetime for our four power law models at $z=2$, 3 and 4. 
As anticipated, the quasar clustering length shows a strong dependence
on quasar lifetime.
The relation between $r_1$ and $\tq$ depends on the power spectrum
index $n$, so the shape of the power spectrum must be known
fairly well to determine $\tq$ from measurements of $r_1$.
The clustering at fixed $\tq/t_U$ is substantially stronger in the
open $n=-1$ model than in the $\om=1$ model because the
underlying mass distribution is more strongly clustered
(larger $\sigma_8$ and $D(z)$).

\begin{figure*}[t]
\centerline{
\epsfxsize=3.5truein
\epsfbox[65 165 550 730]{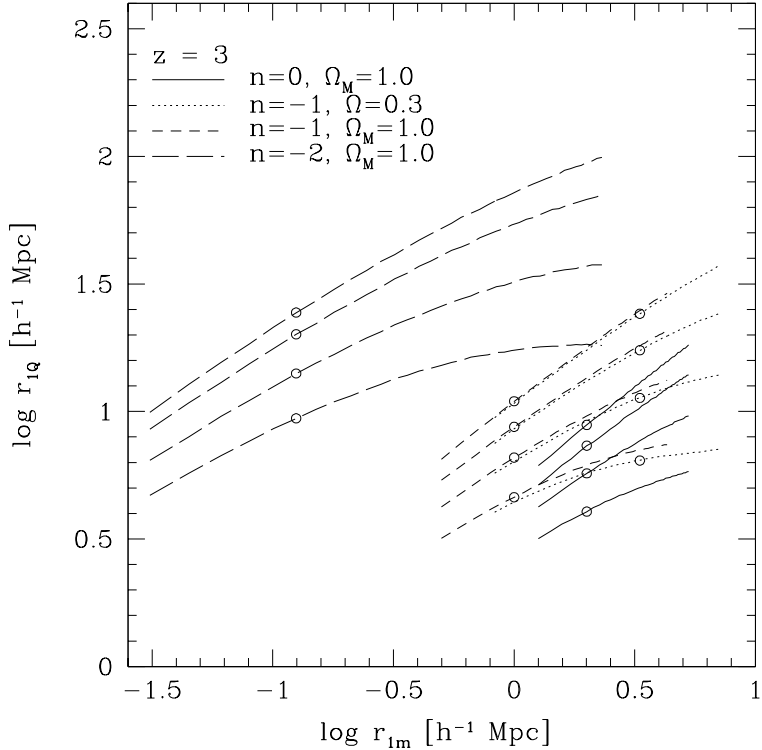}
}
\caption{
\footnotesize
The dependence of the quasar clustering length $r_{1Q}$ on the mass 
clustering length $r_{1m}$, for the four power law models at $z = 3$. 
In each case, the four lines show, from top to bottom, the lifetimes
$\tq = \tu$, $0.1\tu$, $0.01\tu$, and $0.001\tu$.  
Results are computed for normalizations running from $\sigma_8 = 0.2$ to 
$\sigma_8 = 2.0$.  Open circles show 
$r_{1m}$ and $r_{1Q}$ for our standard choices of $\sigma_8$, 
listed in Table~\ref{tbl:pl}.  If bias were independent of $r_{1m}$, the 
lines would parallel the diagonal of the box, which has a slope of $1.0$. 
} \label{fig:r1m}
\end{figure*}

For a specified value of $\om$, the cluster mass function imposes
a reasonably tight constraint on the normalization $\sigma_8$.
It is nonetheless interesting to explore the sensitivity of the
predicted quasar clustering to this normalization.  More intuitive
than the $\sigma_8$-dependence is the equivalent relation between
the quasar clustering length and the corresponding clustering length
of the underlying mass distribution at the same redshift,
\begin{equation}
\ronem = 8\hmpc\,\times\, \left[\sigma_8 D(z)\right]^{2/(3+n)}.
\label{eqn:r1m}
\end{equation}
Figure~\ref{fig:r1m} plots this relation at $z=3$ for the four power
law cosmologies and $\tq = \tu$ (top curve), $0.1\tu$, $0.01\tu$, 
and $0.001\tu$ (bottom curve), for
values of $\sigma_8$ ranging from $0.2$ to $2.0$.
The values of $\ronem$ that correspond to the $\sigma_8$ 
values in Table~\ref{tbl:pl} are marked with open circles. 
If the bias did not change with $\ronem$, then the quasar clustering
length $r_1$ would grow in proportion to $\ronem$, and the curves
in Figure~\ref{fig:r1m} would parallel the diagonal of the box,
which has a slope of 1.0.  However, increasing $\ronem$ increases
$M_*$ and therefore requires a lower value of $\mmin/M_*$ to match
the quasar space density.  The correspondingly lower bias partially
compensates for the larger $\ronem$, making the curves in Figure~\ref{fig:r1m}
shallower than the box diagonal.  For $n=0$ and large $\tq/t_U$ (the
highest solid curve), the minimum mass $\mmin$ lies far out on the tail
of a steeply falling mass function.  In this regime, a change in $M_*$
requires only a small change in $\mmin/M_*$ to compensate, so there 
is little change in $\beff$ with $\ronem$, and the curves approach the
$r_1 \propto \ronem$ lines that would apply for constant bias.
A similar argument explains the steepening of all curves towards 
low $\ronem$, where the small values of $M_*$ put the value of $\mmin$
further out on the tail of the mass function.

\section{Results for Cold Dark Matter (CDM) Cosmologies} \label{sec:cdm}

The results of \S\ref{sec:plspec} confirm our initial contention
that quasar clustering can provide a good diagnostic of the typical
quasar lifetime.  However, they show that the predicted clustering
length also depends on the shape of the mass power spectrum and on
the value of $\om$, which influences the cluster normalization of
$\sigma_8$ at $z=0$ and (together with $\ol$) determines the growth
factor $D(z)$.  Accurate determination of $\tq$ from measurements
of quasar clustering therefore requires reasonably good knowledge
of the underlying cosmology.
Fortunately, many lines of evidence now point towards a flat,
low-density model based on inflation and cold dark matter 
\citep[see, e.g., the review by][]{bahcall99}.  In particular, 
recent studies of the power spectrum of the Ly$\alpha$ forest
imply that the matter power spectrum has the shape and amplitude
predicted by COBE- and cluster-normalized CDM models with $\om \sim 0.4$
at the redshifts and length scales relevant to the prediction
of quasar clustering \citep{croft99,weinberg99,mcdonald00,phillips00}.

For the power spectrum of our CDM models, we adopt 
$P(k) \propto k^{n_p} \, T^2(k)$ with scale-invariant ($n_p=1$)
primeval inflationary fluctuations and the transfer function parameterization
of \citet{bbks86},
\begin{equation} 
T(k) = \frac{ {\rm ln}(1 + 2.34 q)}{2.34 q} \left[ 1 + 3.89 q + (16.1 q)^2 + 
       (5.46 q)^3 + (6.71 q)^4 \right]^{-1/4} .
\label{eqn:trans}
\end{equation}

\noindent
Here $q = k/\Gamma$ and $\Gamma$, with units of $(\hmpc)^{-1}$,
is the CDM shape parameter, given approximately by
$\Gamma = \om h \exp(-\ob-\sqrt{2h}\ob/\om)$ \citep{sugiyama95}.
We calculate $\sigma(M)$ and $(d\sigma/dM)$ by numerical integration
of this power spectrum.

We consider five different CDM models with the parameters 
listed in Table~\ref{tbl:cdm}.  These models are chosen to illustrate
a range of cosmological inputs and also to isolate the effects of 
different parameters on quasar clustering predictions.
The $\tau$CDM, OCDM, and $\Lambda$CDM models have $\Gamma=0.2$,
in approximate agreement with the shape parameter estimated from
galaxy surveys \citep[e.g.,][]{baugh93,peacock94}, 
and they have $\sigma_8$
values consistent with the cluster mass function constraints of 
\citet{eke96}.  The $\tau$CDM and $\Lambda$CDM models are 
approximately COBE-normalized.  COBE normalization would imply
a lower $\sigma_8$ for OCDM, but a slight increase in $n_p$ could
raise $\sigma_8$ without having a large impact on the shape of
$P(k)$ at the relevant scales.  The OCDM and $\Lambda$CDM models
are consistent with the Ly$\alpha$ forest power spectrum measurements
of \citet{croft99}, but the $\tau$CDM model is not.
OCDM is inconsistent with the observed location of the first
acoustic peak in the cosmic microwave background anisotropy
spectrum \citep[e.g.,][]{miller99,melchiorri99,tegmark00}, 
and of the three models, only $\Lambda$CDM is consistent
with the Hubble diagram of Type Ia supernovae \citep{riess98,perlmutter99}.

\begin{center}
\begin{deluxetable}{lccccclll}
\tablenum{2}
\tablewidth{0pt}
\tablecaption{CDM Model Parameters \label{tbl:cdm} }
\tablehead {
  \colhead{} &
  \colhead{$\sigma_8$} &
  \colhead{$h$} &
  \colhead{$\om$} &
  \colhead{$\ol$} &
  \colhead{$\Gamma$} &
  \colhead{$M_* (z = 2)$} &
  \colhead{$M_* (z = 3)$} &
  \colhead{$M_* (z = 4)$} \\
}
\startdata
SCDM          & 0.5  & 0.5  & 1.0 & 0.0 & 0.5 & $3.58\cdot 10^{9}$
  & $2.26\cdot 10^8$    & $1.86\cdot 10^7$    \\
$\tau$CDM     & 0.5  & 0.5  & 1.0 & 0.0 & 0.2 & $9.44\cdot 10^6$
  & $1.09\cdot 10^5$    & $1.87\cdot 10^3$    \\
OCDM          & 0.9  & 0.65 & 0.3 & 0.0 & 0.2 & $1.50\cdot 10^{11}$
  & $2.70\cdot 10^{10}$ & $5.60\cdot 10^{9}$ \\
$\Lambda$CDM  & 0.9  & 0.65 & 0.3 & 0.7 & 0.2 & $2.70\cdot 10^{10}$
  & $2.03\cdot 10^9$    & $1.91\cdot 10^8$    \\
$\Lambda$CDM2 & 1.17 & 0.65 & 0.3 & 0.7 & 0.2 & $2.16\cdot 10^{11}$
  & $2.34\cdot 10^{10}$ & $3.10\cdot 10^9$    \\
\enddata
\tablecomments{Parameters of the five CDM models discussed in \S\ref{sec:cdm}. 
Column 1 lists the model name, columns 2 - 5 the power spectrum normalization 
and cosmological parameters, and column 6 the power spectrum shape 
parameter (see eq.~[\ref{eqn:trans}]). 
Columns 7 - 9 list the values of $M_*$ 
(eq.~[\ref{eqn:mstar}]) at $z = 2$, 3, and 4, in units of $h^{-1} M_\odot$. 
}
\end{deluxetable}
\end{center}

We will use the comparison between the $\tau$CDM and SCDM models,
with $\Gamma=0.2$ and $\Gamma=0.5$, respectively, to illustrate the 
impact of power spectrum shape for fixed $\om$ and $\sigma_8$.
The SCDM model is cluster-normalized, but its $\sigma_8=0.5$
is well below the value $\sigma_8 \approx 1.2$ implied by COBE
for $n_p=1$, $\Gamma=0.5$ \citep[e.g.,][]{bunn97}.
The OCDM and $\Lambda$CDM models have the same $P(k)$ shape and
the same $P(k)$ amplitude at $z=0$, but at high redshift the OCDM
model has stronger fluctuations because of a larger $D(z)$.
We therefore include the model $\Lambda$CDM2, which has $\sigma_8$
chosen to yield the same power
spectrum amplitude as OCDM at $z=3$.  Differences between OCDM and
$\Lambda$CDM2 isolate the impact of a cosmological constant for 
fixed high-redshift mass clustering.

\begin{figure*}[t]
\centerline{
\epsfxsize=3.5truein
\epsfbox[65 165 550 730]{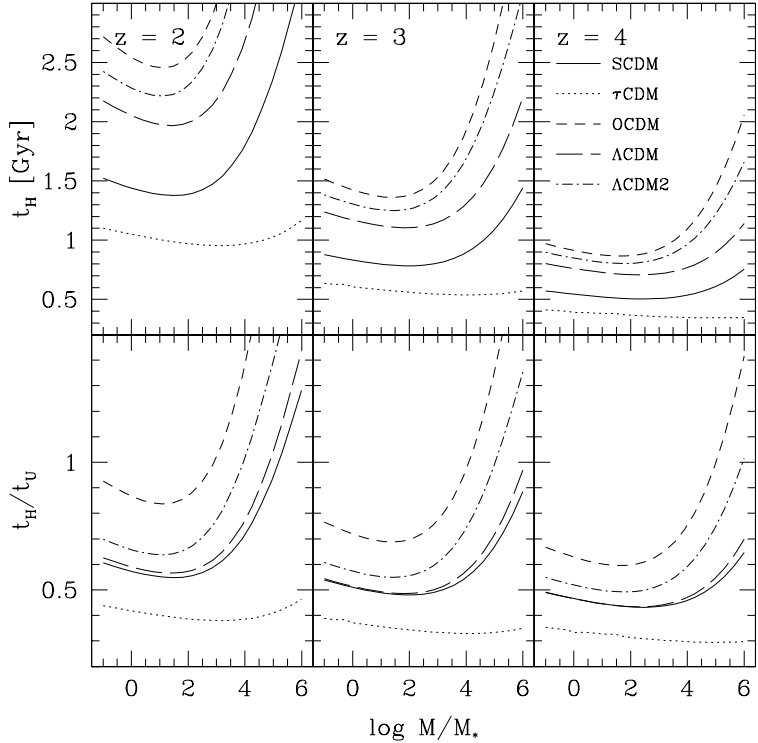}
}
\caption{
\footnotesize
Halo lifetime as a function of $M/M_*$, as in Figure~\ref{fig:lifepl}, 
for the CDM models with parameters listed in Table~\ref{tbl:cdm}. 
} \label{fig:lifecdm}
\end{figure*}

Figure~\ref{fig:lifecdm} shows $t_H$ in Gyr (upper panels) and 
$t_H/\tu$ (lower panels) as a function of $M/M_*$ for the CDM models at
$z = 2$, 3, and 4. 
In contrast to the power law models shown in Figure~\ref{fig:lifepl}, 
the ratio $t_H/t_U$ does not approach a constant value but instead increases
at very large $M/M_*$.  This increase can be understood with reference to 
the power law case: the effective power law index 
$\neff=3-6\,(d\ln\sigma/d\ln M)$ increases
with increasing mass in a CDM spectrum, and larger values of $\neff$
correspond to slower growth of mass scales (and larger $t_H/t_U$) as
shown in Figure~\ref{fig:lifepl}.
The difference between the SCDM and $\tau$CDM curves in 
Figure~\ref{fig:lifecdm} reflects the higher values of $\neff$ 
for the $\Gamma=0.5$ power spectrum.  The differences between the various 
$\Gamma=0.2$ models largely reflect the differences in $M_*$, and
hence the differences in $\neff$ at fixed $M/M_*$, and they also reflect the
differences in fluctuation growth rates.

\begin{figure*}[t]
\centerline{
\epsfxsize=3.5truein
\epsfbox[65 165 550 730]{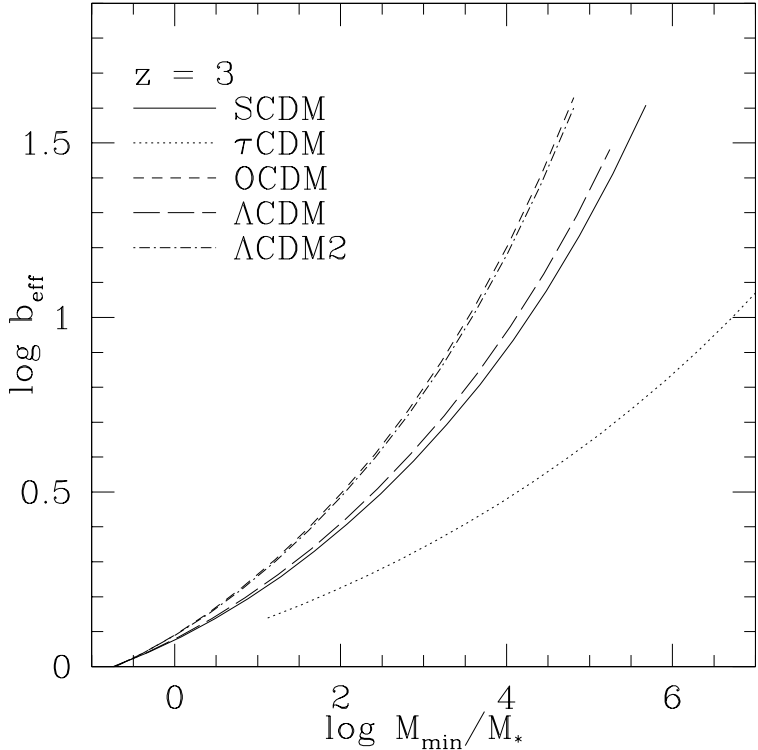}
}
\caption{
\footnotesize
Effective bias as a function of minimum halo mass, as in 
Figure~\ref{fig:biaspl}, for the CDM models at $z = 3$. 
} \label{fig:biascdm}
\end{figure*}

Figure~\ref{fig:biascdm} plots the effective bias against $\mmin/M_*$
for the five CDM models at $z=3$.  Figure~\ref{fig:biaspl} showed
that the value of $\beff$ at fixed $\mmin/M_*$ is higher for larger $n$.
The lines in Figure~\ref{fig:biascdm} curve upwards because $\neff$
increases with mass scale, and to a good approximation the value of $\beff$
in the CDM models equals the value of $\beff$ at the same $\mmin/M_*$
in a power-law model of index $\neff(\mmin)$.
The difference between SCDM and $\tau$CDM in Figure~\ref{fig:biascdm}
therefore reflects the higher $\neff$ values in SCDM, and the 
differences among the other models reflect the different values of $M_*$,
and hence the different values of $\neff$ at fixed $\mmin/M_*$.

\begin{figure*}[t]
\centerline{
\epsfxsize=3.5truein
\epsfbox[65 165 550 730]{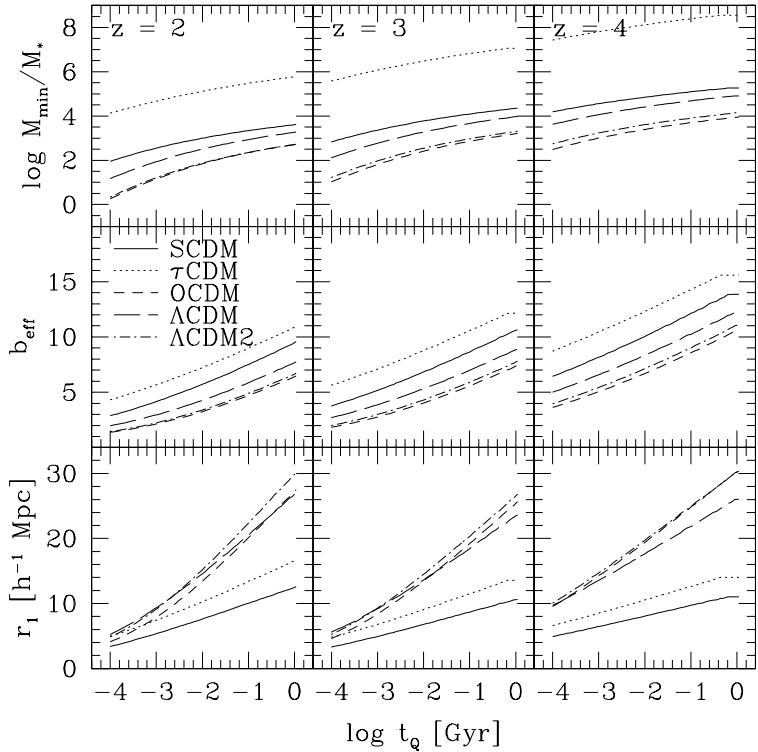}
}
\caption{
\footnotesize
Minimum halo mass, effective bias, and clustering length as a function 
of $\tq$, for the CDM models at $z = 2$, 3, and 4. 
Format is the same as Figure~\ref{fig:r1pl}. 
} \label{fig:r1cdm}
\end{figure*}

The top three panels of Figure~\ref{fig:r1cdm} show the dependence of
$\mmin/M_*$ on $\tq$ at $z = 2$, 3, and 4; the values of $M_*$
are listed in Table~\ref{tbl:cdm}.
The calculation of $\mmin$ via
equation~(\ref{eqn:phimatch2}) incorporates both the dependence of 
halo lifetime on mass and the influence of $\om$ and $\ol$ on the
value of $\Phi(z)$ inferred from the quasar
surface density (as discussed in the Appendix).  
The two $\om = 1$ models have the lowest values of $M_*$
because of their lower $\sigma_8$ and $D(z)$, so they
require the largest $\mmin/M_*$ to match the observed $\Phi(z)$.
The value of $M_*$ is smaller for $\Lambda$CDM than for OCDM because
$D(z)$ is smaller for the flat model, so $\Lambda$CDM requires
larger $\mmin/M_*$.  The higher normalization of the $\Lambda$CDM2
model largely removes this difference, since $\sigma_8 D(z=3)$ is
matched to that of the OCDM model, but $\Lambda$CDM2 still has a 
slightly lower $M_*$ because of the influence of $\Omega_\Lambda$
on $\delta_c(z)$.  As a result, the $\mmin/M_*$ curve for
$\Lambda$CDM2 lies just above that of OCDM at $z=3$.

The middle panels of Figure~\ref{fig:r1cdm} show the effective
bias values, which display the same relative dependence on $\tq$
and cosmology as the $\mmin/M_*$ values.  Because $\mmin/M_* > 1$
in all of the CDM models, even for $\tq$ as low as $10^5$ years,
the MW bias formula~(\ref{eqn:biasmw}) yields nearly identical
results to Jing's (\citeyear{jing98}) corrected formula~(eq.~[\ref{eqn:bias}]).

The bottom panels of Figure~\ref{fig:r1cdm} present the main 
results of this paper: the relation between the clustering length 
$r_1$ and the quasar lifetime $\tq$ for CDM models at $z = 2$, 3, and 4.  
The clustering length is an increasing function of quasar lifetime
for the reasons outlined in the Introduction and detailed in 
\S\ref{sec:meth}.  A longer $\tq$ implies that quasar host halos
are rarer, more highly biased objects.
The change in the $r_1$ vs. $\tq$ relation with redshift 
reflects the evolution of the quasar space density and of the
underlying mass fluctuations.  For a given model and $\tq$,
the predicted quasar clustering is weakest
at $z = 3$, the peak of the quasar space density.
The smaller quasar abundance at $z=4$ implies a higher bias of
the host halo population, which more than compensates for
the slightly weaker mass clustering.
The clustering length grows between $z=3$ and $z=2$ because of
both the drop in quasar space density and the growth of mass clustering.
The $r_1$ vs. $\tq$ relation becomes flat at the largest $\tq$ for the 
SCDM model at $z = 4$ and for the $\tau$CDM model at $z = 3$ and $4$,
where $\tq$ exceeds the halo lifetime $t_H(\mmin)$ and
the value of $\mmin$ required to match $\Phi(z)$ 
therefore becomes independent of $\tq$.

The differences between models reflect the differences in bias
factors discussed above and the differences in the mass clustering.
There are also differences in the values of $\Phi(z)$ inferred
from the observed quasar surface density (see Appendix), but
these have relatively little effect.
The main separation in Figure~\ref{fig:r1cdm} is between 
the low-density models and the $\om=1$ models, which have weaker
mass clustering because of their lower values of $\sigma_8$ and $D(z)$.
The $\om=1$ models have larger bias factors, but these are not enough 
to compensate for the smaller mass fluctuations.
The $r_1$ vs. $\tq$ relations are also shallower for the $\om=1$ models,
because the values of $\mmin$ lie further out on the steep, high-mass tail
of the mass function, where a smaller change in $\mmin$ can make up
for the same change in $\tq$.
The three low-density models yield very similar predictions.

To facilitate comparison of future observational results to our predictions, 
we have fit polynomials of the form
\begin{equation}
r_1 = 	\left\{ \begin{array}{ll}
	 a_0 + a_1 \, \ltq & {\rm for} \; \ltq \geq -1.5 \\
	 a_0 + a_1 \, \ltq + a_2 \, (\ltq + 1.5)^2 & {\rm for} \; \ltq < -1.5\\
        \end{array}
         \right.
\label{eqn:fits}
\end{equation}
to each of the $r_1$ vs. $\tq$ curves shown in Figure~\ref{fig:r1cdm}. 
The values of the coefficients are given in Table~\ref{tbl:fits}, and 
the coefficients $a_0$ and $a_1$ have the same value over the entire 
range in $\tq$.  These fits are accurate to better than 3\% in $r_1$ for 
given $t_Q$, or better than 10\% in $\tq$ given $r_1$, for all cases except 
the SCDM and $\tau$CDM models at $z = 4$, where the maximum errors are 
3\% in $r_1$ given $\tq$ and 20\% in $\tq$ for given $r_1$. 

\begin{center}
\begin{deluxetable}{lccccccccccr}
\tablenum{3}
\tablewidth{0pt}
\tablecaption{$r_1$ vs. $\tq$ fitting coefficients \label{tbl:fits} }
\tablehead {
  \colhead{Model} &
  \colhead{$a_0$} &
  \colhead{$a_1$} &
  \colhead{$a_2$} &
  \colhead{} &
  \colhead{$a_0$} &
  \colhead{$a_1$} &
  \colhead{$a_2$} &
  \colhead{} &
  \colhead{$a_0$} &
  \colhead{$a_1$} &
  \colhead{$a_2$} \\
}
\startdata
              &       &$z = 2$&        & &       &$z=3$&          & &       &$z=4$ &       \\
SCDM          & 12.47 & 2.437 & 0.1001 & & 10.62 & 1.932 & 0.0601 & & 11.26 &1.590 & -0.0042 \\
$\tau$CDM     & 16.54 & 3.163 & 0.1662 & & 13.89 & 2.422 & 0.0882 & & 14.48 &1.964 & -0.0166 \\
OCDM          & 27.18 & 7.002 & 0.7798 & & 25.32 & 6.029 & 0.5347 & & 30.26 &5.457 & 0.1640 \\
$\Lambda$CDM  & 26.79 & 6.173 & 0.4913 & & 23.48 & 4.974 & 0.3122 & & 26.12 &4.248 & 0.0823 \\
$\Lambda$CDM2 & 29.84 & 7.441 & 0.7465 & & 26.51 & 6.093 & 0.4840 & & 30.23 &5.274 & 0.1393 \\
\enddata
\tablecomments{
Coefficients for polynomial fits (eq.~[\ref{eqn:fits}]) to the 
predicted relations between quasar lifetime and clustering length shown in 
Figure~\ref{fig:r1cdm}. Column 1 lists the model name, columns 2 - 4 the 
coefficients for $z=2$, columns 5 - 7 the coefficients for $z=3$, and 
columns 8 - 10 the coefficients for $z=4$. 
}
\end{deluxetable}
\end{center}

\section{Discussion} \label{sec:dis}

\subsection{Sensitivity to model details} \label{sec:sen}

As already mentioned in \S\ref{sec:life}, the definition of a ``halo
lifetime'' is somewhat ambiguous.  We have so far adopted a definition of 
$t_H$ as the median time before a halo of mass $M$ is incorporated
into a halo of mass $2M$.  If we increase this mass ratio from 2 to 5
(a rather extreme value),
then the typical halo lifetimes in our CDM models increase
by factors of $2-4$. 
Since it is the ratio $t_Q/t_H$
that enters our determination of $\mmin$ (eq.~[\ref{eqn:phimatch2}]),
and hence fixes the bias factor, this change in $t_H$ would require
an equal increase in $t_Q$ to maintain the same clustering length $r_1$.
We conclude that the ambiguity in halo lifetime definition 
introduces a factor $\sim 2$ uncertainty in the determination of
$t_Q$ from clustering measurements, in the context of our model.

We have also assumed that quasar luminosity is perfectly correlated
with host halo mass, so that matching the space density of an
absolute-magnitude limited sample imposes a sharp cutoff in 
the host mass distribution at $M=\mmin$.  If there is some scatter
in the luminosity--host mass relation, then some halos with $M<\mmin$
will host a quasar above the absolute-magnitude limit and some
halos with $M>\mmin$ will not.  We can model such an effect
by introducing a soft cutoff into equation~(\ref{eqn:phimatch2}): 
\begin{equation}
\Phi(z) = \int_{0}^{\infty} dM\;g(M)\;\frac{\tq}{t_H(M, z)} n(M, z)
\end{equation}
with
\begin{equation}
g(M) = 	\left\{ \begin{array}{ll}
	0       & {\rm for} \; M < \frac{\mmin}{\alpha} \\
	\left( \frac{\alpha}{\mmin (\alpha^2 - 1)} \right) M - \frac{1}{\alpha^2 - 1}    & {\rm for} \; \frac{\mmin}{\alpha} < M < \alpha \mmin \\
	1       & {\rm for} \; M > \alpha \mmin
        \end{array}
         \right.
\label{eqn:cut}
\end{equation}
and $\alpha > 1$.  Adopting a soft cutoff slightly decreases 
$\mmin$ and, more significantly, reduces the value of $\beff$ by allowing 
some quasars to reside in lower mass halos, which are less strongly biased.
Quantitatively, we find that setting $\alpha = 2$, which corresponds to 
including halos down to $M=\mmin/2$,
decreases the clustering length by $\la 6$\% for the 
shortest quasar lifetimes and $\la 10$\% for the longest quasar 
lifetimes.  Matching a fixed $r_1$ with an $\alpha=2$
cutoff requires lifetimes that are longer by a factor $\sim 1-1.5$
at short $\tq$ and $\sim 2-2.5$ at long $\tq$.
Longer lifetimes are more sensitive to scatter in the luminosity-host mass 
relation because $\beff$ depends more strongly on $\mmin/M_*$ for these 
rarer objects. 
The assumption of a perfectly monotonic relation between quasar
luminosity and host mass leads to the smallest $\tq$ for
a given $r_1$. Thus if any scatter does exist in this relation, our model 
predictions for $\tq$ effectively become lower limits to the quasar 
lifetime. 

Another simplification of our model is the assumption that a quasar is
either ``on'' or ``off'' -- each quasar shines at luminosity $L$ for time
$t_Q$, perhaps divided among several episodes of activity, and the rest
of the time it is too faint to appear in a luminous quasar sample.
More realistically, variations in the accretion rate and
radiative efficiency will cause the quasar luminosity to vary, especially
if the black hole mass itself grows significantly during the 
luminous phase. Nonetheless, the maximum luminosity will still depend on 
the maximum black hole mass. At a given time, the luminous quasar population
will include black holes shining at close to their maximum luminosity and
``faded'' black holes of higher mass. Because the host halos lie on the
steeply falling tail of the mass function, the first component of the
population always dominates over the second, and we therefore expect our
clustering method to yield the time $t_Q$ for which a quasar shines within
a factor $\sim 2$ of its peak luminosity. More strongly faded quasars are too
rare to make much difference to the space density or effective bias.

\begin{figure*}[t]
\centerline{
\epsfxsize=3.5truein
\epsfbox[65 165 550 730]{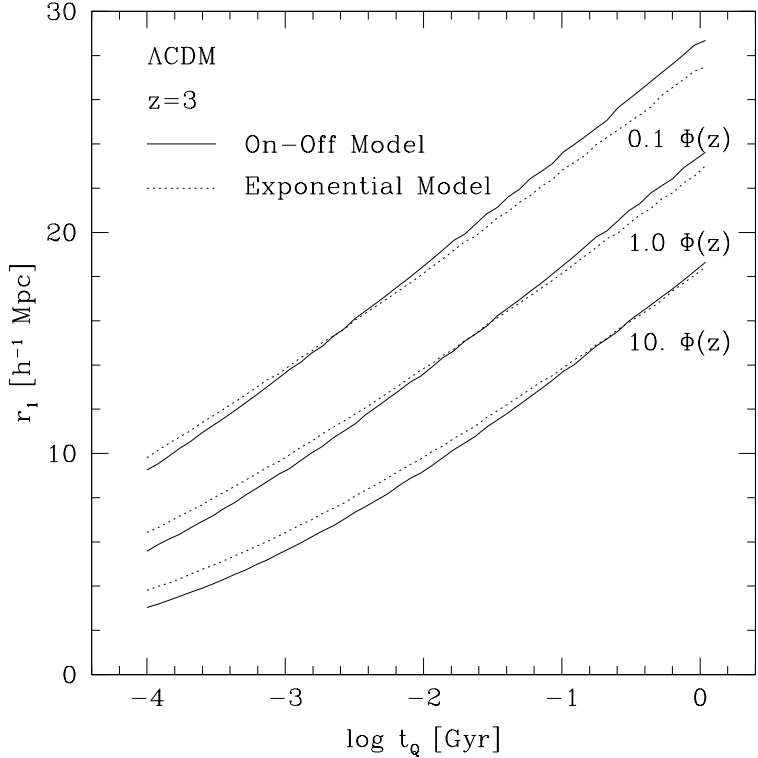}
}
\caption{
\footnotesize
Clustering length vs. $\tq$ for the $\Lambda$CDM model at $z=3$ for 
two different models of quasar luminosity evolution at three different 
space densities $\Phi(z)$. The ``on -- off'' model ({\it solid lines}) 
assumes the quasar luminosity is constant throughout its lifetime $\tq$ 
and is the standard model we discuss in this paper. The central line 
shows results for the $\Lambda$CDM model at $z = 3$ with our 
standard $\Phi(z)$. The other solid curves, related 
to the first by simple horizontal shifts, show results for space densities 
different by factors of $10$ and $1/10$ (bottom and top).
In the exponential model ({\it dotted lines}), the quasar luminosity starts at 
some maximum luminosity proportional to the halo mass and decays with an 
$e$-folding timescale $\tq$. The middle line again corresponds to our 
standard $\Phi(z)$, and the other two dotted curves show results for 
space densities different by factors of $10$ and $1/10$ (bottom and top).
} \label{fig:lt}
\end{figure*}

To illustrate this point, we consider the model of \citet{haehnelt98} in
which a quasar hosted by a halo of mass $M$ has a luminosity history
$L(t) = L_0(M)\,{\rm exp}(-t/t_Q)$, with a maximum luminosity 
$L_0(M) = \alpha\,M$ 
proportional to the halo mass. In this model, the time that a quasar shines 
above the luminosity threshold $L_{\rm min} = L_0 (\mmin)$ of a survey is the
visibility time $t_Q' = t_Q\,{\rm ln}(M/\mmin)$. We can calculate $\mmin$ for a
given space density by substituting $t_Q'$ for $t_Q$ in
equation~(\ref{eqn:phimatch2}), then calculate $\beff(\mmin)$ by
multiplying the integrands in the numerator and denominator of
equation~(\ref{eqn:beff}) by the visibility weighting factor 
${\rm ln}(M/\mmin)$.
The middle curves in Figure~\ref{fig:lt} compare $r_1(t_Q)$ for
the on--off ({\it solid line}) and exponential decay ({\it dotted line})
models, in the case of $\Lambda$CDM at $z = 3$ with our standard $\Phi(z)$.
The curves are remarkably similar, showing that the lifetime inferred from
clustering assuming an on--off model would be close to the $e$-folding
timescale in an exponential decay model. The curves for the exponential
decay model are slightly shallower because at low $\mmin$ (low $t_Q$) the
mass function is not as steep, allowing faded quasars in more massive halos
to make a larger contribution to $\beff$ and thereby raise $r_1$. Although
results for a different functional form of $L(M,t)$ would differ in
detail, we would expect the lifetime inferred from clustering to be
close to the ``half-maximum'' width of the typical luminosity history, for
the general reasons discussed above.

As mentioned in \S\ref{sec:over}, we assume that quasars radiate isotropically. 
If they radiate instead with an average beaming factor $f_B < 1$, then 
the true value of $\Phi(z)$ is larger than the observed value by a 
factor $f_B^{-1}$. The implied lifetime for a given $r_1$ would therefore 
be larger by a factor $f_B^{-1}$ as well.

\subsection{Interpretation of Existing Data} \label{sec:data}

After several attempts \citep{osmer81, webster82}, quasar clustering was 
first detected by \citet{shaver84}, and later by \citet{shanks87} and 
\citet{iovino88}.  However, measurements of quasar clustering are
still hampered by small, sparse samples, and even the best studies
to date yield detections with only several-$\sigma$ significance.
Given the limitations of current data, it is not surprising that
different authors reach different conclusions about the strength of
clustering and its evolution.  Analyzing a combined sample of quasars
with $0.3<z<2.2$ from the Durham/AAT UVX Survey, 
the CFHT survey, and the  Large Bright Quasar Sample,
\citet{shanks94} and \citet{croom96} find a reasonable fit to the data with
an $\om=1$ model that has $\xi(r)=(r/r_0)^{-\gamma}$, $\gamma \approx 1.8$,
and a constant comoving correlation length $r_0=6\hmpc$.
\citet{lafranca98} report a higher correlation length, 
$r_0=9.1 \pm 2.0\hmpc$ for a $\gamma=1.8$ power law, in their
$1.4 < z < 2.2$ sample.

If we adopt $r_0 \approx 8\hmpc$ at $z=2$ and a corresponding
$r_1 \approx 11\hmpc$, then the implied quasar lifetime is 
$\sim 10^{7.5}$ years for the $\tau$CDM model and $\sim 10^8$ years
for SCDM.  The $r_0$ values quoted above are for $\om=1$,
and because quasar pair separations are measured in angle and redshift,
they should be increased by a factor $\sim 1.5$ in an $\om=0.3$, $\ol=0.7$
universe and a factor $\sim 1.3$ in an $\om=0.3$, $\ol=0$ universe
(roughly the inverse cube-roots of the volume ratios listed in 
Table A1).  Adopting $r_1 \approx 16\hmpc$ implies a lifetime
$t_Q \sim 10^{7-7.5}$ years in our low-density models.
However, these numbers must be considered highly uncertain because of the 
limitations of current data and because the space densities of the various 
observational samples do not necessarily match those assumed in our 
model predictions. 

All of these measurements are based
mainly on quasars with $z<2$.  At higher redshift, \citet{kundic97}
and \citet{stephens97} have investigated clustering in the
Palomar Transit Grism Survey (PTGS; \citealt{schneider94}).
Fitting a $\gamma=1.8$ power law, \citet{stephens97} find
$r_0=17.5 \pm 7.5\hmpc$ for $z>2.7$.  This high
correlation length (inferred from the presence of three close
pairs in a sample of 90 quasars) could be a statistical fluke,
but in the context of our model it is tempting to see it as
a consequence of the high luminosity threshold of the PTGS
survey, which might lead it to pick out the most strongly clustered members 
of the quasar population.

\subsection{Prospects} \label{sec:pros}

The 2dF \citep{boyle00,shanks00} and Sloan \citep{york00} quasar surveys
will transform the study of quasar clustering over the next
several years, yielding high-precision measurements for a wide
range of redshifts.  These measurements will allow good determination
of the typical quasar lifetime $\tq$ in the context of the model 
presented here.  They will also test the key assumption of this model, 
the monotonic
relation between quasar luminosity and host halo mass, by
characterizing the clustering as a function of redshift and,
especially, as a function of quasar absolute magnitude.

Figure~\ref{fig:lt} illustrates this test for the $\Lambda$CDM model 
at $z=3$. Brighter quasars have a lower space density $\Phi(z)$,
so they should have a higher minimum host halo mass $\mmin$, and,
because of the higher bias of more massive halos, they should 
exhibit stronger clustering.  Fainter, more numerous quasars
should exhibit weaker clustering.  Figure~\ref{fig:lt} shows 
the predicted $r_1$ vs. $\tq$ relation for samples with 1/10
and 10 times the space density of our standard case (3.42 quasars
per square degree per unit redshift; see Table~\ref{tbl:qlf}).
In our standard on--off model ({\it solid lines}), 
a change in $\Phi(z)$ in equation~(\ref{eqn:phimatch2}) 
can be exactly compensated by changing $\tq$ by the same factor,
so the solid curves in Figure~\ref{fig:lt} are simply shifted horizontally
relative to each other.  Our predictions in 
Figure~\ref{fig:r1cdm} (see eq.~[\ref{eqn:fits}]) can therefore be
transformed to any quasar space density by changing $\tq$ in
proportion to $\Phi(z)$. In the exponential decay model ({\it dotted lines}), 
the scaling of $t_Q$ with $\Phi(z)$ is no longer exact, though it 
is still a good approximation. 

If there is a large dispersion in the relation between quasar luminosity
and host halo mass, then the dependence of clustering strength on
quasar space density will be much weaker than Figure~\ref{fig:lt} predicts.  
Detection of the predicted trend between luminosity and clustering, or 
definitive demonstration of its absence, would itself provide an important
insight into the nature of quasar host halos.
More generally, the parameters of a model that incorporates scatter
(such as the $\alpha$ prescription of equation~[\ref{eqn:cut}])
could be determined by matching the observed relation
between $r_1$ and $\Phi(z)$.

If the observations do support a tight correlation between luminosity
and host halo mass, then the first property of quasars to emerge from 
the 2dF and Sloan clustering studies will be the typical lifetime $\tq$.
For the low-density models in Figure~\ref{fig:r1cdm},
the slope of the correlation between $r_1$ and $\log_{10}\tq$ is $\sim 10$,
so a determination of $r_1$ with a precision of $2\hmpc$ would
constrain $\tq$ to a factor of $10^{0.2} \approx 1.6$, for 
a specified cosmology.  By the time these quasar surveys are complete,
a variety of observations may have constrained cosmological
parameters to the point that they contribute negligible uncertainty
to this constraint.  Instead, the uncertainty in $\tq$ will probably
be dominated by the limitations of the quasar population model,
e.g., the approximate nature of the
assumptions that the quasar luminosity tracks the halo mass,
that there is only quasar per halo, and that the average lifetime $\tq$
is independent of quasar luminosity.
These assumptions can be tested empirically to some degree, but not
perfectly.
Despite these limitations, it seems realistic to hope that 
$\tq$ can be constrained to a factor three or better by high-precision
clustering measurements, a vast improvement over the current situation.
It is worth reiterating that our assumption of a perfectly monotonic
relation between luminosity and halo mass leads to the smallest $\tq$
for an observed $r_1$, since with a shorter lifetime there are simply
not enough massive, highly biased halos to host the quasar population.

A determination of $\tq$ to a factor of three will be sufficient
to address fundamental issues about the physics of quasars and
galactic nuclei.  Comparison of $\tq$ to the Salpeter timescale
will answer one of the most basic questions about supermassive
black holes: do they shine as they grow?
If $\tq \ga 4\times 10^7$ years, the $e$-folding timescale
for $L \sim L_E$, $\epsilon \sim 0.1$, then quasar black holes increase 
their mass by a substantial factor during their optically bright phase.  
If $\tq$ is much shorter than this, then the black holes must accrete most
of their mass at low efficiency, or while shining at $L \ll L_E$.
A short lifetime could indicate an important role for advection 
dominated accretion (\citet{narayan98} and references therein), or it 
could indicate that black holes acquire much of their mass through
mergers with other black holes, emitting binding energy in the form of 
gravitational waves rather than electromagnetic waves.
A determination of $\tq$ would also resolve the question
of whether the black holes in the nuclei of local galaxies
are the remnants of dead quasars.
For example, \citet{richstone98} infer a lifetime $t_Q \sim 10^6$ years by
matching the space density of local
spheroids that host black holes of mass $M \ga 4\times 10^8 M_\odot$ 
to the space density of high-redshift quasars
of luminosity $L_E(M) \ga 6 \times 10^{46}\;{\rm erg}\;{\rm s}^{-1}$.
If clustering implies a much longer lifetime, then these numerous local 
black holes may once have powered active nuclei, but they were not the engines
of the luminous, rare quasars.

We have assumed in our model that quasar activity is a random 
event in the life of the parent halo.
Quasar activity might instead be triggered by a major merger,
by a weaker ``fly-by'' interaction, or by the first burst of
star formation in the host galaxy.
Regardless of the trigger mechanism, the lifetime will be the
dominant factor in determining the strength of high-redshift quasar 
clustering, if our assumed link between luminosity
and halo mass holds.  However, different triggering mechanisms
might be diagnosed by more subtle clustering properties,
such as features in the correlation function at small separations,
or higher-order correlations.  At low redshift, where the evolution
of the quasar population is driven by fueling rather than by
black hole growth, the nature of the triggering mechanism might
play a major role in determining quasars' clustering properties.
The calculations presented here illustrate the promise of quasar
clustering as a tool for testing ideas about quasar physics, a promise 
that should be fulfilled by the large quasar surveys now underway. 

\acknowledgments
We thank James Bullock, Alberto Conti, Jordi Miralda-Escud{\'e}, 
Patrick Osmer, and Simon White for helpful discussions. 
We also thank the referee for constructive suggestions, which led to 
our consideration of the exponential decay model in \S~\ref{sec:dis}. 
As we were nearing completion of this work, we learned of a
similar, independent study by Z. Haiman \& L. Hui 
\citep{haiman00}; our general conclusions are consistent with theirs, 
although the approaches are quite different in detail, precluding a 
precise comparison of results. 
This work was supported in part by NSF grant AST96-16822 and
NASA grant NAG5-3525.

\appendix

\section{Converting from observed quasar numbers to $\Phi(z)$} 

The observed quantity that is measured in studies of quasar clustering and 
the quasar space density is the number of sources brighter than a given 
apparent magnitude $m$ per unit redshift per unit solid angle on the sky. 
This surface density per unit redshift can be converted into a comoving space 
density of objects brighter than a given absolute magnitude $M$, 
\begin{equation}
\Phi(z, < M) = \frac{dN (< m)}{d\Omega \; dz} \; \frac{d\Omega \; dz}{dV_c(z)}, 
\end{equation}
where $dV_c(z)$ is the differential comoving volume element corresponding to 
$d\Omega \, dz$.  
Following the notation in \citet{hogg99}, this volume element is 
\begin{equation}
dV_c(z) = D_H \frac{D_M^2}{E(z)} d\Omega dz, 
\end{equation}
\noindent
where $D_H = c/H_0$ is the Hubble distance, $D_M$ is the transverse 
comoving distance, 
\begin{equation}
D_M = 	\left\{ \begin{array}{ll}
	D_H \, \frac{1}{\sqrt{\Omega_k}} \, {\rm sinh}[\sqrt{\Omega_k}\frac{D_c}{D_H} ] & {\rm for} \; \Omega_k > 0 \\
	D_H \int_{0}^{z} \frac{dz'}{E(z')} & {\rm for} \; \Omega_k = 0 \\
	D_H \frac{1}{\sqrt{|\Omega_k}|} \, {\rm sin} [\sqrt{|\Omega_k|}\frac{D_c}{D_H} ] & {\rm for} \; \Omega_k < 0 
	\end{array}
	\right.
\end{equation}
\noindent 
and 
$E(z) = \sqrt{\om(1+z)^3 + \Omega_k(1+z)^2 + \ol}$,  
where $\Omega_k = 1 - \om - \ol$. 
For $\om = 1$, $\Omega_k = 0$, the differential comoving volume element is 
\begin{equation}
dV_C(z) = 4 \, \left(\frac{c}{H_0}\right)^3 \; (1 + z)^{-3/2} \; \left[ 1 - \frac{1}{\sqrt{1 + z}} \right]^2 \, d\Omega \, dz
\end{equation}
per steradian per unit redshift. 

The fact that $dV_c(z)$ depends on the cosmological parameters means that 
a given measured surface density of sources corresponds to a different 
comoving space density for different cosmological model parameters. 
The space density of quasars is commonly quoted for an $\om = 1$ universe.  
To convert this space density (in units of $h^3$ Mpc$^{-3}$) into the space 
density for a model with different values of $\om$ and $\ol$ requires a 
correction of the form
\begin{equation} 
\Phi(z, \om, \ol) = 
	\Phi(z, \om', \ol') 
	\frac{f(z, \om', \ol')}
	{f(z, \om, \ol)}, 
\end{equation}
where
\begin{equation} 
f(z, \om, \ol) = \frac{D_H \, D_M^2}{E(z)}. 
\label{eqn:f}
\end{equation}
This procedure converts the reported space density under one set of 
cosmological parameters back into the observed surface density and 
then converts the surface density into the space density for the new 
set of cosmological parameters. In the notation of \citet{pwro98}, 
$f(z, \om, \ol) = g \times f^2$, where $f$ and $g$ are given by their 
equations (5) and (6), respectively. 

\begin{center}
\begin{deluxetable}{lrcccc}
\tablenum{A1}
\tablewidth{0pt}
\tablecaption{Quasar Space and Surface Density \label{tbl:qlf} }
\tablehead {
  \colhead{$z$} &
  \colhead{} &
  \colhead{$\Phi(z)$} &
  \colhead{$\frac{dN}{dz\,d\Omega}$} &
  \colhead{$\frac{f(1.0, 0.0)}{f(0.3, 0.0)}$} &
  \colhead{$\frac{f(1.0, 0.0)}{f(0.3, 0.7)}$} \\
}
\startdata
2 & . . . . . . . . . . . . . . . . & $1.889\cdot 10^{-6}$ & 2.132 & 0.425 & 0.279 \\
3 & . . . . . . . . . . . . . . . . & $3.331\cdot 10^{-6}$ & 3.417 & 0.339 & 0.253 \\
4 & . . . . . . . . . . . . . . . . & $3.200\cdot 10^{-7}$ & 0.287 & 0.287 & 0.241 \\
\enddata
\tablecomments{Adopted values of the space density of quasars at 
$z = 2$, 3, and 4, 
and cosmological conversion factors.  Values of $\Phi(z)$ in column 2 are from 
\citet{who94} (assuming $\om = 1$) for quasars with $M_c < -24.5$ (absolute 
continuum flux at 1216 \AA), converted from their adopted $h = 0.75$ to 
$h^3$ Mpc$^{-3}$. 
Column 3 lists the abundance of quasars in number per square 
degree per unit redshift to which the space density in column 2 
corresponds. Columns 4 and 
5 contain the ratios of the factors $f(\om, \ol)$ defined in 
equation~(\ref{eqn:f}) needed to convert the space density in column 2, 
which is valid for $\om = 1$, to space densities for the OCDM and 
$\Lambda$CDM models, respectively. 
}
\end{deluxetable}
\end{center}

In Table~\ref{tbl:qlf} we list the factors to convert the space density 
in column 2, which is listed for $\om = 1, \ol = 0$, 
to the corresponding space densities for $\om = 0.3, \ol = 0.0$ and 
$\om = 0.3, \ol = 0.7$. The factors in 
Table~\ref{tbl:qlf} are all less than unity because the comoving volume 
element is smallest in an $\om = 1$ universe. 

{}


\begin{thebibliography}{}

\bibitem[Adelberger et al.(1998)]{adelberger98}
Adelberger, K. L., Steidel, C. C., Giavalisco, M., Dickinson, M.,
Pettini, M., \& Kellogg, M. 1998, \apj, 505, 18

\bibitem[Bahcall et al.(1999)]{bahcall99}
Bahcall, N. A., Ostriker, J. P., Perlmutter, S., \& Steinhardt, P. J. 1999,
Science, 284, 1481

\bibitem[Baugh \& Efstathiou(1993)]{baugh93}
Baugh, C. M., \& Efstathiou, G. 1993, \mnras, 265, 145

\bibitem[Bajtlik, Duncan, \& Ostriker(1988)]{bajtlik88}
Bajtlik, S., Duncan, R.C., \& Ostriker, J.P. 1988, \apj, 327, 570

\bibitem[Bardeen et al.(1986)]{bbks86} 
Bardeen, J., Bond, J.R., Kaiser, N., \& Szalay, A.S. 1986, \apj, 304, 15

\bibitem[Bechtold(1994)]{bechtold94}
Bechtold, J., 1994, \apjs, 91, 1

\bibitem[Blumenthal et al.(1984)]{blumenthal84}
Blumenthal, G.R., Faber, S.M., Primack, J.R., \& Rees, M.J. 1984, 
\nat, 311, 517

\bibitem[Bond et al.(1991)]{bond91}
Bond, J. R., Cole, S., Efstathiou, G., \& Kaiser, N. 1991, \apj, 379, 440

\bibitem[Boyle et al.(1990)]{boyle90} 
Boyle, B.J., Fong, R., Shanks, T., \& Peterson, B.A. 1990, \mnras, 243, 1

\bibitem[Boyle et al.(1999)]{boyle99}
Boyle, B. J., Smith, R.
J., Shanks, T., Croom, S. M. \& Miller, L. 1999, in IAU Symp. 183:
Cosmological Parameters and the Evolution of the Universe, ed. K. Sato,
(Kluwer: Dordrecht), p. 178

\bibitem[Boyle et al.(2000)]{boyle00}
Boyle, B.J., Shanks, T., Croom, S.M., Smith, R.J., Miller, L., Loaring, N., 
\& Heymans, C. 2000, \mnras, {\it in press} astro-ph/0005368

\bibitem[Bunn \& White(1997)]{bunn97}
Bunn, E. F., \& White, M. 1997, \apj, 480, 6

\bibitem[Carroll, Press, \& Turner(1992)]{carroll92}
Carroll, S. M., Press, W. H., \& Turner, E. L. 1992, \araa, 30, 499

\bibitem[Ciotti \& Ostriker(1999)]{ciotti99}
Ciotti, L., \& Ostriker, J. P. 1999, \apj, {\it submitted}, astro-ph/9912064

\bibitem[Croft et al.(1999)]{croft99}
Croft, R. A. C., Weinberg, D. H., Pettini, M., Katz, N., \& Hernquist, L. 1999,
\apj, 520, 1

\bibitem[Croom \& Shanks(1996)]{croom96} 
Croom, S.M. \& Shanks, T. 1996, \mnras, 281, 893

\bibitem[Efstathiou \& Rees(1988)]{efstathiou88}
Efstathiou, G. \& Rees, M. J. 1988, \mnras, 230, 5P

\bibitem[Eke, Cole, \& Frenk(1996)]{eke96}
Eke, V. R., Cole, S., \& Frenk, C. S. 1996, \mnras, 282, 263

\bibitem[Faber \& Jackson(1976)]{faber76}
Faber, S. M. \& Jackson, R. E. 1976, \apj, 204, 668

\bibitem[Fan et al.(1999)]{fan99}
Fan, X., et al.\ 1999, \aj, 118, 1

\bibitem[Haehnelt \& Rees(1993)]{haehnelt93} 
Haehnelt, M.G. \& Rees, M.J. 1993, \mnras, 263, 168

\bibitem[Haehnelt, Natarajan, \& Rees(1998)]{haehnelt98}
Haehnelt, M. G., Natarajan, P., \& Rees, M. J. 1998, \mnras, 300, 817

\bibitem[Haiman \& Hui(2000)]{haiman00}
Haiman, Z.  \& Hui, L. 2000, \apj, submitted, astro-ph/0002190

\bibitem[Haiman \& Loeb(1998)]{haiman98}
Haiman, Z.  \& Loeb, A.  1998, \apj, 503, 505

\bibitem[Heath(1977)]{heath77}
Heath, D. J. 1977, \mnras, 179, 351

\bibitem[Hewett, Foltz, \& Chaffee(1993)]{hewett93}
Hewett, P.C., Foltz, C.B., \& Chaffee, F.C. 1993, \apj, 406, L43

\bibitem[Hogg(1999)]{hogg99} 
Hogg, D.W. 1999, astro-ph/9905116

\bibitem[Iovino \& Shaver(1988)]{iovino88}
Iovino, A. \& Shaver, P.A. 1988, \apj, 330, L13

\bibitem[Jing(1998)]{jing98}
Jing, Y.P. 1998, \apj, 503, L9

\bibitem[Kaiser(1984)]{kaiser84}
Kaiser, N. 1984, \apj, 294, L9

\bibitem[Katz, Hernquist, \& Weinberg(1999)]{katz99}
Katz, N., Hernquist, L., \& Weinberg, D. H. 1999, \apj, 523, 463

\bibitem[Katz et al.(1994)]{katz94}
Katz, N., Quinn, T., Bertschinger, E., \& Gelb, J. M. 1994,
\mnras, 270, L71

\bibitem[Kauffmann \& Haehnelt(2000)]{kauffmann00}
Kauffmann, G., \& Haehnelt, M. 2000, \mnras, 311, 576

\bibitem[Kolatt et al.(1999)]{kolatt99}
Kolatt, T. S., et al. 1999, \apjl, 523, L109 

\bibitem[Kundi\'c(1997)]{kundic97} 
Kundi\'c, T. 1997, \apj, 482, 631

\bibitem[La Franca et al.(1998)]{lafranca98}
La Franca, F., Andreani, P., \& Cristiani, S. 1998, \apj, 497, 529

\bibitem[Lacey \& Cole(1993)]{lc93} 
Lacey, C. \& Cole, S. 1993, \mnras, 262, 649 

\bibitem[Lynden-Bell(1969)]{lyndenbell69}
Lynden-Bell, D. 1969, Nature, 223, 690

\bibitem[Magorrian et al.(1998)]{mag98} 
Magorrian, J., et al. 1998, \aj, 115, 2285

\bibitem[McDonald et al.(2000)]{mcdonald00}
McDonald, P., Miralda-Escud\'e, J., Rauch, M., Sargent, W. L. W.,
Barlow, T. A., Cen, R., \& Ostriker, J. P. 2000, \apj, submitted,
astro-ph/9911196

\bibitem[Melchiorri et al.(1999)]{melchiorri99}
Melchiorri, A. et al.\ 1999, \apjl, in press, astro-ph/9911445

\bibitem[Miller et al.(1999)]{miller99}
Miller, A. D. et al.\ 1999, \apj, 524, L1

\bibitem[Mo \& White(1996)]{mw96} 
Mo, H.J. \& White, S.D.M. 1996, \mnras, 282, 347 (MW)

\bibitem[Mo, Mao, \& White(1999)]{mmw99} 
Mo, H.J., Mao, S., \& White, S.D.M. 1999, \mnras, 304, 175

\bibitem[Monaco, Salucci, \& Danese(2000)]{monaco00}
Monaco, P., Salucci, P., \& Danese, L. 2000, \mnras, 311, 279

\bibitem[Narayan, Mahadevan, \& Quatert(1998)]{narayan98}
Narayan, R., Mahadevan, R., \& Quatert, E. 1998, in 
The Theory of Black Hole Accretion Discs, 
ed. M. A. Abramowicz, G. Bjornsson, \& J. E. Pringle, 
(Cambridge University Press), 148

\bibitem[Navarro, Frenk, \& White(1997)]{nfw97} 
Navarro, J.F., Frenk, C.S., \& White, S.D.M. 1997, \apj, 490, 493

\bibitem[Osmer(1981)]{osmer81} 
Osmer, P.S. 1981, \apj, 247, 762

\bibitem[Osmer(1998)]{osmer98}
Osmer, P. S. 1998, in 
ASP Conference Series 146, The Young Universe,
ed. S. D'odorico, A. Fontana, E. Giallongo, (San Francisco: ASP), 1

\bibitem[Peacock \& Dodds(1994)]{peacock94}
Peacock, J. A., \& Dodds, S. J. 1994, \mnras, 267, 1020

\bibitem[Peebles(1980)]{peebles80}
Peebles, P. J. E. 1980, The Large Scale Structure of the
Universe (Princeton: Princeton University Press)

\bibitem[Perlmutter et al.(1999)]{perlmutter99}
Perlmutter, S., et al. 1999, \apj, 517, 565

\bibitem[Phillips et al.(2000)]{phillips00}
Phillips, J., Croft, R. A. C., Weinberg, D.H., Hernquist, L., Katz, N., \&
Pettini, M. 2000, \apj, submitted, astro-ph/0001089

\bibitem[Press \& Schechter(1974)]{ps74} 
Press, W.H. \& Schechter, P. 1974, \apj, 187, 425 (PS)

\bibitem[Popowski et al.(1998)]{pwro98} 
Popowski, P.A., Weinberg, D.H., Ryden, B.S., \& Osmer, P.S. 1998, \apj, 498, 11

\bibitem[Richstone et al.(1998)]{richstone98}
Richstone, D., et al.\ 1998, Nature, 395 Supp, A14

\bibitem[Riess et al.(1998)]{riess98}
Riess, A. G., et al.\ 1998, \aj, 116, 1009

\bibitem[Salucci et al.(1999)]{salucci99}
Salucci, P., Szuszkiewicz, E., Monaco, P., \& Danese, L. 1999, \mnras,
307, 637

\bibitem[Salpeter(1964)]{salpeter64}
Salpeter, E. E. 1964, \apj, 140, 796

\bibitem[Schneider, Schmidt, \& Gunn(1994)]{schneider94}
Schneider, D. P., Schmidt, M., \& Gunn, J. E. 1994, \aj, 107, 1245

\bibitem[Shanks \& Boyle(1994)]{shanks94} 
Shanks, T. \& Boyle, B.J. 1994, \mnras, 271, 753

\bibitem[Shanks et al.(1987)]{shanks87}
Shanks, T., Fong, R., Boyle, B.J., \& Peterson, B.A. 1987, \mnras, 227, 739

\bibitem[Shanks et al.(2000)]{shanks00}
Shanks, T., Boyle, B.J., Croom, S.M., Loaring, N., Miller, L., Smith, R.J.
2000, in Clustering at High Redshift, eds. A. Mazure, O. LeFevre, V. Lebrun, 
{\it in press}, astro-ph/0003206

\bibitem[Shaver(1984)]{shaver84}
Shaver, P.A. 1984, \aap, 136, L9

\bibitem[Stephens et al.(1997)]{stephens97} 
Stephens, A.W., Schneider, D.P., Schmidt, M., Gunn, J.E., \& Weinberg, D.H. 
1997, \aj, 114, 41

\bibitem[Sugiyama(1995)]{sugiyama95}
Sugiyama, N. 1995, \apjs, 100, 281

\bibitem[Tegmark \& Zaldarriaga(2000)]{tegmark00}
Tegmark, M. \& Zaldarriaga, M. 2000, \apj, {\it in press}, astro-ph/0002091

\bibitem[Turner(1991)]{turner91}
Turner, E. L. 1991, \aj, 101, 5

\bibitem[van der Marel(1999)]{marel99}
van der Marel, R. P. 1999, \aj, 117, 744

\bibitem[Warren, Hewett, \& Osmer(1994)]{who94} 
Warren, S.J., Hewett, P.C., \& Osmer, P.S. 1994, \apj, 421, 412

\bibitem[Weinberg et al.(1999)]{weinberg99}
Weinberg, D. H., Croft, R. A. C., Hernquist, L., Katz, N., \& Pettini, M. 1999,
\apj, 522, 563

\bibitem[White, Efstathiou, \& Frenk(1993)]{white93}
White, S. D. M., Efstathiou, G. P., \& Frenk, C. S. 1993, \mnras, 262, 1023

\bibitem[Webster(1982)]{webster82}
Webster, A. 1982, \mnras, 199, 683

\bibitem[York et al.(2000)]{york00}
York, D., et al. 2000, \aj, {\it in press}

\bibitem[Zel'dovich \& Novikov(1964)]{zeldovich64}
Zel'dovich, Ya. B. \& Novikov, I.D. 1964, Sov. Phys. Dokl., 158, 811

\end{thebibliography}
\end{document}